\documentclass{aastex63}

\usepackage{color}
\usepackage{graphicx}

\received{}
\revised{}
\accepted{}
\submitjournal{ApJ}

\shorttitle{ALMA observations of dense cores in Taurus}
\shortauthors{Tokuda et al.}

\begin{document}

\title{FRagmentation and Evolution of dense cores Judged by ALMA (FREJA). I (Overview).  Inner $\sim$1000\,au structures of prestellar/protostellar cores in Taurus}

\correspondingauthor{Kazuki Tokuda}
\email{tokuda@p.s.osakafu-u.ac.jp}
\author[0000-0002-2062-1600]{Kazuki Tokuda}
\affiliation{Department of Physical Science, Graduate School of Science, Osaka Prefecture University, 1-1 Gakuen-cho, Naka-ku, Sakai, Osaka 599-8531, Japan}
\affiliation{National Astronomical Observatory of Japan, National Institutes of Natural Science, 2-21-1 Osawa, Mitaka, Tokyo 181-8588, Japan}

\author{Kakeru Fujishiro}
\affiliation{Department of Physics, Nagoya University, Chikusa-ku, Nagoya 464-8602, Japan}

\author{Kengo Tachihara}
\affiliation{Department of Physics, Nagoya University, Chikusa-ku, Nagoya 464-8602, Japan}

\author{Tatsuyuki Takashima}
\affiliation{Department of Physical Science, Graduate School of Science, Osaka Prefecture University, 1-1 Gakuen-cho, Naka-ku, Sakai, Osaka 599-8531, Japan}

\author{Yasuo Fukui}
\affiliation{Department of Physics, Nagoya University, Chikusa-ku, Nagoya 464-8602, Japan}
\affiliation{Institute for Advanced Research, Nagoya University, Furo-cho, Chikusa-ku, Nagoya 464-8601, Japan}

\author{Sarolta Zahorecz}
\affiliation{Department of Physical Science, Graduate School of Science, Osaka Prefecture University, 1-1 Gakuen-cho, Naka-ku, Sakai, Osaka 599-8531, Japan}
\affiliation{National Astronomical Observatory of Japan, National Institutes of Natural Science, 2-21-1 Osawa, Mitaka, Tokyo 181-8588, Japan}

\author{Kazuya Saigo}
\affiliation{National Astronomical Observatory of Japan, National Institutes of Natural Science, 2-21-1 Osawa, Mitaka, Tokyo 181-8588, Japan}

\author{Tomoaki Matsumoto}
\affiliation{Faculty of Sustainability Studies, Hosei University, Fujimi, Chiyoda-ku, Tokyo 102-8160, Japan}

\author{Kengo Tomida}
\affiliation{Astronomical Institute, Tohoku University, 6-3, Aramaki, Aoba-ku, Sendai, Miyagi 980-8578, Japan}

\author{Masahiro N. Machida}
\affiliation{Department of Earth and Planetary Sciences, Faculty of Sciences, Kyushu University, Nishi-ku, Fukuoka 819-0395, Japan}

\author{Shu-ichiro Inutsuka}
\affiliation{Department of Physics, Nagoya University, Chikusa-ku, Nagoya 464-8602, Japan}

\author{Philippe Andr\'e}
\affiliation{Laboratoire d'Astrophysique (AIM), CEA, CNRS, Universit\'e Paris-Saclay, Universit\'e Paris Diderot, Sorbonne Paris Cit\'e, 91191 Gif-sur-Yvette, France}

\author{Akiko Kawamura}
\affiliation{National Astronomical Observatory of Japan, National Institutes of Natural Science, 2-21-1 Osawa, Mitaka, Tokyo 181-8588, Japan}

\author[0000-0001-7826-3837]{Toshikazu Onishi}
\affiliation{Department of Physical Science, Graduate School of Science, Osaka Prefecture University, 1-1 Gakuen-cho, Naka-ku, Sakai, Osaka 599-8531, Japan}

\begin{abstract}

We have performed survey-type observations in 1\,mm continuum and molecular lines toward dense cores (32 prestellar + 7 protostellar) with an average density of $\gtrsim$10$^5$\,cm$^{-3}$ in the Taurus molecular clouds using the Atacama Large Millimeter/submillimeter Array–Atacama Compact Array (ALMA-ACA) stand-alone mode with an angular resolution of 6\farcs5 ($\sim$900\,au). The primary purpose of this study is to investigate the innermost part of dense cores with view to understanding the initial condition of star formation. 
In the protostellar cores, contributions from protostellar disks dominate the observed continuum flux with a range of 35\%--90\% except for the very low-luminosity object. For the prestellar cores, we have successfully confirmed continuum emission from dense gas with a density of $\gtrsim$3\,$\times$\,10$^5$\,cm$^{-3}$ toward approximately one-third of the targets. Thanks to the lower spatial frequency coverage with the ACA 7\,m array, the detection rate is significantly higher than that of the previous surveys, which have zero or one continuum-detected sources among a large number of starless samples using the ALMA Main Array.
The statistical counting method tells us that the lifetime of the prestellar cores until protostar formation therein approaches the freefall time as the density increases. Among the prestellar cores, at least two targets have possible internal substructures, which are detected in continuum emission with the size scale of $\sim$1000\,au if we consider the molecular line (C$^{18}$O and N$_2$D$^{+}$) distributions. These results suggest that small-scale fragmentation/coalescence processes occur in a region smaller than 0.1\,pc, which may determine the final core mass associated with individual protostar formation before starting the dynamical collapse of the core with central density of $\sim$(0.3--1)\,$\times$\,10$^6$\,cm$^{-3}$.

\end{abstract}


\section{Introduction} \label{sec:intro}

Understanding of protostar formation via the gravitational collapse of a single molecular cloud core (hereafter dense core; e.g., \citealt{Myers83}) is one of the most fundamental issues in astrophysics because the stars are the minimum ingredients in galaxies.
In addition, protostellar disks as the residual product of the cloud collapse \citep{Terebey84} eventually form planets \citep{Dutrey14}. Although a large number of theoretical and observational works in the last few decades established an overall picture of low-mass star formation \citep[see the review by][]{Shu87}, many fundamental questions remain to be studied. In particular, understanding the origin of the initial mass function (IMF) is considered to be one of the ultimate goals of star formation studies. The mass spectrum of dense cores may reflect the mechanism of their formation/fragmentation and should be relevant to the IMF \citep{Goodwin08}. 

\subsection{Single-dish Surveys toward Dense Cores in Nearby Star-forming Regions} \label{intro:single-dish}
Single-dish surveys using bolometer cameras revealed that the core mass function (CMF) in the cluster-forming regions of $\rho$\,Ophiuchus resembles the stellar IMF, although their observations lack sufficient samples above 0.5\,$M_\odot$ \citep{Motte98}. The subsequent unbiased survey using a high-density ($\sim$10$^5$\,cm$^{-3}$) molecular gas tracer confirmed similar results through their compilation of a much robust number of the sample \citep{Onishi96,Onishi98,Onishi02,Tachihara02}. Deep submillimeter continuum maps \citep[e.g.,][]{Nutter07} found that the CMF turns over at $\sim$1\,$M_\odot$, and the slope of the low-mass side is also consistent with that of the IMF \citep[see also][]{Konyves15,Marsh16}.
Such large-scale surveys indicate that the core-to-star formation efficiency is $\sim$20--40\% \citep[e.g.,][]{Alves07,Andre10} assuming that the CMF is converted to the IMF by the one-to-one transformation. This efficiency is consistent with that realized by the mass loss due to protostellar outflow, as expected from numerical simulations \citep[e.g.,][]{Machida12}. Such a dense parental core prior to protostar formation, called a prestellar cores, is supposed to be gravitationally bound (formerly, pre-protostellar cores; see \citealt{Ward-Thompson94}). Note that recent surveys suggested that some starless sources are likely pressure confined \cite[e.g.,][]{Pattle15,Kirk17b}. Although the above-mentioned relation between the CMF and IMF may not be simple (see the review by \citealt{Offner14}),
the current observational indications tell us that, likely, the nature of prestellar cores may reflect the properties of forming stars, such as stellar mass, and the multiplicity inside them. It is thus important to characterize their properties as the initial condition for star formation \citep[for reviews, e.g.,][]{Bergin07,Ward-Thompson07}.
Note that such surveys are also vital in the search for rare objects, such as very high-density cores just before/after star formation whose timescale is very short \citep{Mizuno94,Onishi99}.

Early millimeter/submillimeter continuum emission and near-infrared extinction studies toward individual dense cores without any indications of star formation demonstrated that most of the starless cores have inner flat density structures \citep[e.g.,][]{Ward-Thompson94} and the (column) density profiles in Bok globules are well characterized by a critical Bonnor--Ebert (BE) sphere \citep{Ebert55,Bonnor56} model \citep{Alves01,Kandori05}. Numerical simulations for the gravitational collapse of a dense core often adopt such a relatively simple geometry as an initial condition \citep[e.g.,][]{Machida08}. However, \cite{Kirk05} showed that the BE model could not be applied to one-third of the observed dense cores and suggested that they are already collapsing or supported by an additional force, such as the magnetic field. \cite{Onishi02} found a large number of irregular-shaped cores with an average density of $\sim$10$^5$\,cm$^{-3}$ in Taurus. Although their tracer, H$^{13}$CO$^+$, possibly traces lower density gas and affected by the molecular depletion \citep[see][]{Caselli02a}, the observations of N$_2$H$^+$ and its deuterated species, which are less sensitive to this problem, also show similar results for some cores \citep[e.g.,][]{Tafalla15,Punanova18}. Although virial analysis of these objects shows that they are gravitationally bound objects, the presence of such irregular structures imply that environmental effects such as turbulence are still not negligible in the core dynamics. Numerical simulation by \cite{Padoan02} claimed that the inner density profile with two-dimensional averaging mimics a BE-like structure even if the core has an irregular shape created by turbulent compression.  In any case, the previous single-dish observational studies did not have sufficient angular resolution to resolve the inner $\sim$1000\,au regions of prestellar cores. Therefore, they characterized them as uniform density parts.

\subsection{Interferometric Observations of Prestellar Cores} \label{intro:array}
Recent observational studies in the early phase of star formation (i.e., class 0/I phases) suggest that most of the stars can be formed as binary/multiple members \citep[e.g.,][]{Chen13} whose frequency is $\sim$2/3. The classic idea to explain such a system is fragmentation in gravitationally unstable massive disks \citep{Larson87,Boss02,Machida08} after the protostar formation. Some numerical simulations adopting strong turbulence produce local high-density maxima, possibly leading to binary/multiple star formation within a single core \citep[e.g.,][]{Offner10}. \cite{Tobin15,Tobin16} suggest that such a turbulent-driven fragmentation can be a promising candidate as the origin of wide binary/multiple systems based on their systematic survey, and \cite{Pineda15} found filamentary clouds, possibly leading to a quadruple star system with a separation of $\gtrsim$1000\,au. 

Recent ALMA observations of the protostellar core MC27/L1521F in Taurus resolved multiple overdense peaks, which cannot be explained by the coherent collapsing motion alone \citep{Tokuda14,Tokuda16,Tokuda17}. Some ALMA studies \citep[e.g.,][]{Williams14,Fern17,Lee17} found binary/multiple protostellar disks whose projected rotational axes are highly misaligned with each other, possibly driven by turbulence. It is crucial to search for fragmented structures at the prestellar collapse phase and investigate their physical properties when we consider the origin of such complex systems.
Nevertheless, taking into account the fact that most of the prestellar cores have a flat density profile in the early phase of cloud collapse as suggested by theoretical studies \citep[e.g.,][]{Larson69} and observations (see Sect.\,\ref{intro:single-dish}), detecting compact emission is a challenging task due to the spatial filtering effect of interferometers. 

Early interferometric observations studied millimeter continuum emission \citep{Schnee10,Schnee12,Andre12,Nakamura12,Friesen14,Ohashi18,Tatematsu20} and found hidden protostellar objects including the first hydrostatic core (hereafter FHSC, e.g., \citealt{Larson69,Masunaga98,Tomida13}) candidate \citep[e.g.,][]{Chen10,Chen12,Enoch10,Pineda11,Pezzuto12,Hirano14,Karnath20} toward infrared-quiescent sources, especially in nearby cluster-forming regions, such as Perseus and $\rho$ Ophiuchus \citep[see the recent ALMA survey by][]{Kirk17}. However, these studies cannot clarify how an isolated dense core evolves into protostar(s) without contaminations from surrounding phenomena (e.g., stellar feedback). Investigations toward low-mass star-forming complexes, such as Taurus and Bok Globules, are vital to understanding the process of dense core evolution and possible fragmentation leading to a binary/multiple system inside a single core \citep[see also the introduction in][]{Caselli19}. The fact that the intrinsic column densities of Taurus dense cores are one order of magnitude lower than those in Perseus \citep[Figure 7 in][]{Ward-Thompson07} makes it more difficult for us to find compact/evolved features, which are detectable with interferometers. As one of the most prominent examples, \cite{Dunham16} could not find any 3\,mm continuum emission inside 56 starless sources in Chamaeleon\,I using the ALMA Main array (12\,m array). More recently, \cite{Caselli19} found a high-density ($\sim$10$^7$\,cm$^{-3}$) compact peak toward one of the most well-studied prestellar cores, L1544, in Taurus. 
The critical next step is to reveal universality/diversity regarding the prestellar evolution in a molecular complex, for example, the presence/absence of fragmentation within a single core and their evolution timescale by the statistical counting. 

The previous study by \cite{Dunham16} was a pilot survey to search for extremely high-density objects whose densities are $\gtrsim$10$^{8}$\,cm$^{-3}$, possibly in the FHSC stage. \cite{Onishi02} implied that the threshold density for the dynamical collapse of cores is $\sim$10$^{6}$\,cm$^{-3}$, and thus investigations of the properties of prestellar cores with a density of 10$^{5}$--10$^{6}$\,cm$^{-3}$ is crucial in order to understand the condition for the onset of star formation. The Jeans length ($\lambda_{\rm J}$ = $\sqrt{\pi c_{\rm s}/G\rho_{\rm 0}}$, where $G$ is the gravitational constant, $\rho_{\rm 0}$ is the mean density, and $c_{\rm s}$ is the sound speed of 0.2\,km\,s$^{-1}$ at 10\,K) of 10$^{6}$\,cm$^{-3}$ gas is $\sim$2000\,au, which corresponds to $\sim$14\arcsec at nearby star-forming regions, such as Taurus, Ophiuchus-North, Lupus, and Chamaeleon, whose distances are $\sim$150\,pc \citep[e.g.,][]{Schlafly14}. Such a scale is similar to the beam size of the single-dish telescopes, and the ALMA 12\,m array may fail to detect the relatively extended emission due to the filtering-out effect \citep{Onishi13}. 
The Atacama Compact Array (ACA, a.k.a Morita Array), which includes short-spacing baselines, is the best tool to explore the innermost part of a dense core without a serious missing flux problem compared to the 12\,m array alone. We demonstrated the capability of the ACA by discovering a candidate of the prestellar core, possibly leading to brown dwarf or very low-mass star formation \citep{Tokuda19}. The work was the pilot study of the present project.

We present a survey in 1.3\,mm continuum and molecular lines toward 39 dense cores (7 protostellar + 32 prestellar cores) in Taurus using the ACA. We name this project as $``$FRagmentation and Evolution of dense cores Judged by ALMA (FREJA),$"$ which is composed of multiple ALMA campaigns (see Sect.\,\ref{sec:obs}). Our primary strategy is (1) to find evolved features inside prestellar cores and perform statistical analysis to constrain theories of dense core evolution using the ACA stand-alone mode, and (2) to carry out follow-up observations with the 12m array to further characterize the initial condition of low-mass star formation. In this paper (Paper I), we describe the survey design of this project with the ACA stand-alone mode and especially the early results revealed by the 1.3 mm continuum observations. \citet[hereafter Paper~II]{Fujishiro20} investigate the nature of a possible FHSC candidate in L1535NE/MC35.

\section{Observations and Data Reduction} \label{sec:obs}
\subsection{Survey Design and Descriptions of the Observations} \label{obs:design}
\subsubsection{Target Selection}
We observed a large number of dense cores (32 prestellar core + 7 protostellar cores in total) in Taurus to investigate their inner structures with the ACA stand-alone mode in ALMA Cycles 4 and 6. Although our primary targets are prestellar cores, we included protostellar cores associated with young stellar objects (YSOs) for comparison purposes. In the Cycle 4 program (P.I., K. Tachihara, \#2016.1.00928.S, hereafter PROJ4), we targeted 16 dense cores located at several nearby ($D\sim$150\,pc) low-mass, star-forming regions---Taurus, Ophiuchus-North, Chamaeleon, and Lupus---based on the available continuum data obtained by single-dish telescopes \citep[see the description in][]{Tokuda19}. Because we have confirmed the $\sim$50\% detection rate in the millimeter continuum emission from prestellar sources, we planned to observe more objects to search for evolved sources and perform statistical analyses.

To obtain a large number that is as unbiased as possible, we targeted the Taurus molecular cloud as the observation region in the Cycle 6 program (P.I., K. Tachihara, \#2018.1.00756.S, hereafter PROJ6). The Nagoya 4\,m telescope with $\sim$3\arcmin\,resolution performed an unbiased survey in $^{13}$CO\,($J$ = 1--0) across the Taurus complex \citep{Mizuno95}, and the subsequent C$^{18}$O\,($J$ = 1--0) observations revealed distributions of dense ($\sim$10$^4$\,cm$^{-3}$) filamentary molecular clouds \citep{Onishi96,Onishi98}. Based on the large-scale mapping, high-resolution H$^{13}$CO$^{+}$\,(1--0) observations using the Nobeyama 45\,m telescope \citep{Onishi02} discovered 44 prestellar cores with a density of $\gtrsim$10$^5$\,cm$^{-3}$ toward high-column density ($>$10$^{22}$\,cm$^{-2}$) regions in Taurus. Such cores detectable in high-density gas tracers are gravitationally bound, and thus they are considered to be prestellar cores (\citealt{Ward-Thompson94,Ward-Thompson07}; see also \citealt{Marsh14} for the Taurus L1495 region). H$^{13}$CO$^+$ sometimes cannot trace the column density peak, possibly due to the depletion effect onto dust grains in cold/dense environments \cite[e.g.,][]{Caselli02a}. Because this problem happens over a single-dish beam-size-scale even in an optically thin case, the confirmation of another molecular line tracer, such as a N-bearing species, is used to select evolved prestellar cores robustly.
We investigated the N$_2$H$^{+}$ detection toward the \cite{Onishi02} catalog in the literature \citep{Tatematsu04,Tafalla04,Tafalla06,Tafalla15,Punanova18} and our independent measurements obtained with the Nobeyama 45\,m telescope \citep[Y. Miyamoto et al. in preparation; see also][]{Tokuda19}. Note that the recent large-scale surveys in the B213/L1495 and B18 regions with the Green Bank Telescope \citep{Seo15,Seo19,Friesen17} also detected NH$_3$ emission toward our selected samples. 

\begin{figure}[htbp]
\includegraphics[width=180mm]{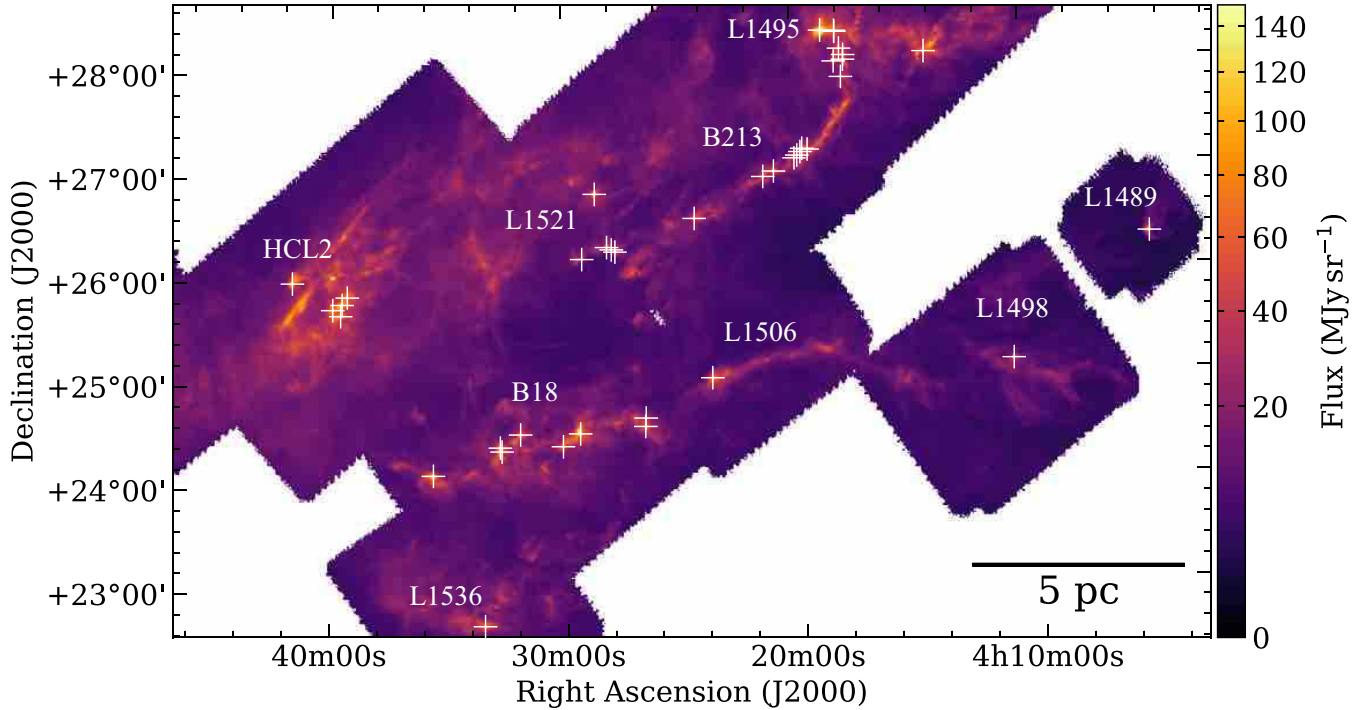}
\caption{Observation targets in the ACA survey on the Herschel/SPIRE 500\,$\micron$ continuum map \cite[e.g.,][]{Palmeirim13}. The white crosses denote the central coordinates of the observed field.  
\label{fig:Taurusmap}}
\end{figure}

We selected the observation coordinates to the peak positions of the  millimeter/submillimeter continuum emission obtained by the IRAM\,30\,m, the JCMT\,15\,m, and the Herschel telescope \citep{Kauffmann08,Palmeirim13,Marsh16,Ward-Thompson16,Tokuda19}. We set the selection criteria in H$_2$ column density as $\gtrsim$10$^{22}$\,cm$^{-2}$, which is higher than the column density threshold for prestellar cores \citep{Andre10}. 
If there are two local peaks in the continuum emission on a single core cataloged by \cite{Onishi02}, we targeted two fields to observe each peak. These targets are MC5N/S, MC7N/S, MC13a/W, MC25E/W, and MC33bN/S. Note that there are two targets \citep[B10 and L1521E; see][]{Hirota02,Hacar13,Tafalla15} that were not included in the \cite{Onishi02} catalog. The total number of observed fields is 32 as prestellar sources. We included seven protostellar cores, whose evolutionary stages are mostly class 0/I phases, to investigate differences between pre-/protostellar cores. We previously performed a case study toward the MC27/L1521F class\,0 Very Low-Luminosity Object (VeLLO) system using a similar frequency setting \citep{Tokuda17,Tokuda18}. We include the data set (hereafter, PROJ-MC27) in this paper. 
Table\,\ref{tab:target} and Figure\,\ref{fig:Taurusmap} give the properties of the targeted objects and their positions on the large-scale Herschel dust continuum image, respectively. 

\begin{table}[htbp]
\begin{center}
\caption{Target objects}
\label{tab:target}
\begin{tabular}{lccccccccc} \hline
Name & RA & Dec & $n_{\rm ave}$(H$_2$)$^{\rm a}$ & $T_{\rm k}^{\rm b}$ & Distance$^{\rm c}$ & Stage$^{\rm d}$ & Other name & Region & Project\\ 
     & (J2000.0) & (J2000.0) & ($10^5\,\rm{cm}^{-3}$) & (K) & (pc) & & & & \\ \hline
MC1 & 4$^{\rm h}$04$^{\rm m}$48\fs0 & +26\arcdeg19\arcmin22\farcs0 & $\cdots$ & 10.0 & 126.6 & Pre & L1489-NH$_3$ $^{\rm e,f}$& L1489 & PROJ6 \\
MC2 & 4$^{\rm h}$10$^{\rm m}$52\fs3 & +25\arcdeg10\arcmin04\farcs0 & 0.8 & $\cdots$  & 126.6 & Pre & L1498$^{\rm e}$ & L1498 & PROJ6 \\
MC4 & 4$^{\rm h}$14$^{\rm m}$11\fs0 & +28\arcdeg09\arcmin12\farcs0 & 1.1 & $\cdots$  & 129.5 & Pre & L1495SE$^{\rm e}$ & L1495 & PROJ6 \\
MC5N & 4$^{\rm h}$17$^{\rm m}$42\fs0 & +28\arcdeg08\arcmin43\farcs0 & 1.9 & 9.3 & 129.5 & Pre & $\cdots$ & L1495 & PROJ4,PROJ6 \\
MC5S & 4$^{\rm h}$17$^{\rm m}$43\fs0 & +28\arcdeg06\arcmin01\farcs0 & 1.9 & 9.7 & 129.5 & Pre & $\cdots$ & L1495 & PROJ6 \\
MC6 & 4$^{\rm h}$17$^{\rm m}$53\fs3 & +28\arcdeg13\arcmin15\farcs0 & 1.2 & 9.5 & 129.5 & Pre & $\cdots$ & L1495 & PROJ4 \\
MC7N & 4$^{\rm h}$18$^{\rm m}$03\fs5 & +28\arcdeg22\arcmin57\farcs0 & 1.9 & 9.7 & 129.5 & Pre & $\cdots$ & L1495 & PROJ6 \\
MC7S & 4$^{\rm h}$18$^{\rm m}$02\fs9 & +28\arcdeg22\arcmin18\farcs0 & 1.9 & 9.7 & 129.5 & Pre & $\cdots$ & L1495 & PROJ6 \\
MC8 & 4$^{\rm h}$18$^{\rm m}$07\fs5 & +28\arcdeg05\arcmin14\farcs0 & 1.0  & 9.3 & 129.5 & Pre & $\cdots$ & L1495 & PROJ4 \\
MC11 & 4$^{\rm h}$18$^{\rm m}$39\fs9 & +28\arcdeg23\arcmin18\farcs0 & 0.5 & 10.3 & 129.5 & Pre & $\cdots$ & L1495 & PROJ6 \\
B10& 4$^{\rm h}$17$^{\rm m}$50\fs0 & +27\arcdeg56\arcmin07\farcs0 & $\cdots$ & 9.3 & 129.5 & Pre & $\cdots$ & L1495 & PROJ6 \\
MC13W & 4$^{\rm h}$19$^{\rm m}$23\fs7 & +27\arcdeg14\arcmin50\farcs5 & 1.4 & 9.2 & 158.1 & Pre & $\cdots$ & B213 & PROJ6 \\
MC13a & 4$^{\rm h}$19$^{\rm m}$37\fs5 & +27\arcdeg15\arcmin23\farcs0 & 1.4 & 9.7 & 158.1 & Pre & $\cdots$ & B213 & PROJ6 \\
MC13b & 4$^{\rm h}$19$^{\rm m}$42\fs5 & +27\arcdeg13\arcmin32\farcs0 & 3.1 & 10.4 & 158.1 & Class\,0/I & $\cdots$ & B213 & PROJ6 \\
MC14N & 4$^{\rm h}$19$^{\rm m}$51\fs0 & +27\arcdeg11\arcmin34\farcs0 & $\cdots$ & 9.1 &  158.1 & Pre & 04169-NW$^{\rm f}$ & B213 & PROJ6 \\
MC14S & 4$^{\rm h}$19$^{\rm m}$58\fs5 & +27\arcdeg11\arcmin59\farcs5 & 5.0 & 10.3 & 158.1 & Class\,0/I & $\cdots$ & B213 & PROJ6 \\
MC16E & 4$^{\rm h}$21$^{\rm m}$21\fs0 & +26\arcdeg59\arcmin47\farcs0 & 2.5 & 9.5 & 158.1 & Pre & L1521D$^{\rm g}$ & B213 & PROJ6 \\
MC16W & 4$^{\rm h}$20$^{\rm m}$53\fs0 & +27\arcdeg02\arcmin49\farcs0 & 1.8 & 9.9 & 158.1 & Pre & L1521D$^{\rm g}$ & B213 & PROJ6 \\
MC19 & 4$^{\rm h}$23$^{\rm m}$43\fs5 & +25\arcdeg04\arcmin13\farcs0 & 0.7  & $\cdots$ & 126.6 & Pre & $\cdots$ & L1506 & PROJ6 \\
MC22 & 4$^{\rm h}$24$^{\rm m}$21\fs6 & +26\arcdeg36\arcmin34\farcs0 & 2.1 & $\cdots$ & 140 & Pre & L1521B$^{\rm e}$ & L1521 & PROJ6 \\
MC23 & 4$^{\rm h}$26$^{\rm m}$37\fs0 & +24\arcdeg36\arcmin52\farcs5 & 0.7 & $\cdots$ & 126.6 & Pre & $\cdots$ & B18 & PROJ6 \\
MC24 & 4$^{\rm h}$26$^{\rm m}$35\fs5 & +24\arcdeg41\arcmin44\farcs0 & 2.4 & 10.7 & 126.6 & Pre & $\cdots$ & B18 & PROJ6 \\
MC25E & 4$^{\rm h}$28$^{\rm m}$09\fs5 & +26\arcdeg20\arcmin43\farcs5 & 1.2 & $\cdots$ & 140 & Pre & B217$^{\rm e}$ & L1521 & PROJ6 \\
MC25W & 4$^{\rm h}$27$^{\rm m}$48\fs0 & +26\arcdeg18\arcmin02\farcs0 & 1.2 & $\cdots$ & 140 & Pre & B217$^{\rm e}$ & L1521 & PROJ6 \\
MC26a & 4$^{\rm h}$27$^{\rm m}$57\fs4 & +26\arcdeg19\arcmin19\farcs0 & $\cdots$ & $\cdots$ & 140 & Class\,I & $\cdots$ & L1521 & PROJ6 \\
MC27 & 4$^{\rm h}$28$^{\rm m}$39\fs0 & +26\arcdeg51\arcmin39\farcs9 & 1.6 & 9.1 & 140 & VeLLO & L1521F$^{\rm g}$ & L1521 & PROJ-MC27 \\
L1521E & 4$^{\rm h}$29$^{\rm m}$14\fs0 & +26\arcdeg14\arcmin00\farcs0 & $\cdots$ & $\cdots$ & 140 & Pre & $\cdots$ & L1521 & PROJ6\\
MC28 & 4$^{\rm h}$29$^{\rm m}$22\fs9 & +24\arcdeg33\arcmin09\farcs5 & 1.3 & 11.0 & 126.6 & Class\,I & $\cdots$ & B18 & PROJ6 \\
MC29 & 4$^{\rm h}$30$^{\rm m}$07\fs0 & +24\arcdeg25\arcmin52\farcs0 & 0.8 & 9.4 & 126.6 & Pre & $\cdots$ & B18 & PROJ6 \\
MC31 & 4$^{\rm h}$31$^{\rm m}$55\fs5 & +24\arcdeg32\arcmin55\farcs0 & 1.2 & 8.6 & 126.6 & Pre & TMC-2A$^{\rm e}$ & B18 & PROJ4,PROJ6 \\
MC33bS & 4$^{\rm h}$32$^{\rm m}$43\fs1 & +24\arcdeg23\arcmin09\farcs0 & 1.5 & 9.2 & 126.6 & Pre & TMC-2$^{\rm h}$& B18 & PROJ4,PROJ6 \\
MC33bN & 4$^{\rm h}$32$^{\rm m}$47\fs3 & +24\arcdeg25\arcmin26\farcs6 & 1.5 & 9.1 & 126.6 & Pre & TMC-2$^{\rm h}$& B18 & PROJ6 \\
MC34 & 4$^{\rm h}$33$^{\rm m}$27\fs7 & +22\arcdeg42\arcmin09\farcs1 & 0.6 & $\cdots$ & 162.7 & Pre & L1536$^{\rm e}$ & L1536 & PROJ6 \\
MC35 & 4$^{\rm h}$35$^{\rm m}$37\fs5 & +24\arcdeg09\arcmin17\farcs0 & 4.0 & 10.0  & 126.6 & Pre & L1535NE$^{\rm i}$ & B18 & PROJ6 \\
MC37 & 4$^{\rm h}$39$^{\rm m}$17\fs8 & +25\arcdeg52\arcmin23\farcs1 & 1.1 & 11.2 & 137.0 & Pre & $\cdots$ & HCL2 & PROJ6 \\
MC38 & 4$^{\rm h}$39$^{\rm m}$31\fs0 & +25\arcdeg48\arcmin05\farcs0 & 0.6 & $\cdots$ & 137.0 & Pre & $\cdots$ & HCL2 & PROJ6 \\
MC39 & 4$^{\rm h}$39$^{\rm m}$34\fs7 & +25\arcdeg41\arcmin34\farcs0 & 1.0 & $\cdots$ & 137.0 & Class\,I & $\cdots$ & HCL2 & PROJ6 \\
MC41 & 4$^{\rm h}$39$^{\rm m}$55\fs0 & +25\arcdeg45\arcmin03\farcs0 & 1.1 & $\cdots$ & 137.0 & Class\,II & $\cdots$ & HCL2 & PROJ6 \\
MC44 & 4$^{\rm h}$41$^{\rm m}$39\fs2 & +26\arcdeg00\arcmin15\farcs0 & 0.7 & 9.9 & 137.0 & Pre & TMC-1C$^{\rm e}$ & HCL2 & PROJ4 \\
\hline
\end{tabular}\end{center}
$^{\rm a}$Average number H$_2$ density estimated from H$^{13}$CO$^{+}$ observations \citep{Onishi02}.
$^{\rm b}$Kinematic temperature derived from NH$_3$ observations \citep{Benson89,Codella97,Seo15,Feh16,Friesen17}.
$^{\rm c}$Distance measurements toward individual clouds in Taurus by \cite{Galli18}. Because there is no available data for the L1521 region in their study, we applied 140\,pc \citep[e.g.,][]{Elias78} for it.
$^{\rm d}``$Pre$"$ means prestellar cores. The other sources are protostellar cores containing VeLLO \citep{Bourke06}, class\,0/I \citep{Motte01}, class\,I \citep{Kenyon93a,Kenyon93b}, and class II \citep{Hill12}.
$^{\rm e-i}$\cite{Benson89}, \cite{Motte01}, \cite{Codella97}, \cite{Myers79}, \cite{Hogerheijde00}, respectively.
\end{table}

\subsubsection{Frequency Setting} \label{obs:freq}
Millimeter/submillimeter continuum emission is a fundamental tracer of the column density distribution of molecular clouds because the emission is optically thin under the conditions of a typical interstellar medium. According to the ALMA sensitivity calculator, Band\,6 (211--275\,GHz) is one of the most suitable frequencies to maximize the signal-to-noise ratio of thermal dust emission assuming the spectral index, $\alpha$, is more than 2. For the 7\,m array observations in Band\,6, the angular resolution of $\sim$6\arcsec, which is a factor of 2 or 3 higher than that of previous single-dish observations, and the $\sim$30\arcsec maximum recovering scale allow us to distinguish whether of not the innermost part has a smoothed column density. 

In addition to the continuum observations, molecular lines are essential for tracing gas kinematics of the high-density part. In PROJ4, we included two molecular lines, H$^{13}$CO$^{+}$\,(3--2) and H$^{13}$CN\,(3--2), as the dense gas tracers. However, the ACA 7\,m array observations could not find significant emission of either line above the 3$\sigma$ level toward all the six Taurus targets (see Table\,\ref{tab:target}) even in the three cases (MC5N, MC31, and MC33bN) where the dust emission was detected. We thus changed our strategy in PROJ6. The target lines were N$_2$D$^{+}$\,(3--2), $^{12}$CO\,(2--1), $^{13}$CO\,(2--1), and C$^{18}$O\,(2--1). Although some observational studies toward evolved prestellar cores, such as L183 and L1544, show indications of molecular depletion in N$_2$D$^{+}$ at the core center \citep{Pagani07,Redaelli19}, the line basically correlates well with the dust continuum in cold ($\sim$10K) and high-density (10$^5$--10$^6$\,cm$^{-3}$) environments. 

Because the CO isotopologs show highly extended distributions across molecular clouds and often become optically thick toward dense regions, they are not always suitable as a dense gas tracer. However, recent high-resolution observations suggest that they have some benefits in investigating gas properties in an early phase of star formation. For example, if there are embedded protostars or FHSC candidates within the cores, the $^{12}$CO observations can trace the outflow activity (e.g., \citealt{Pineda11}, see also Paper II). The optically thick $^{12}$CO lines also work like a thermometer. \cite{Tokuda18} found warm (15-60 K) CO gas components possibly generated by a turbulent shock in a cold protostellar core, MC27/L1521F, in Taurus. This type of observation may provide us with a new method to witness a moment of turbulent dissipation at an early phase of star formation \citep{Tachihara00,Pon12,Pon14,Larson15}. The remaining CO isotopologs, $^{13}$CO and C$^{18}$O, fill a gap in the traced density range between N$_2$D$^{+}$ and $^{12}$CO. In the MC27/L1521F study \citep{Tokuda18}, $^{13}$CO and C$^{18}$O traced a peculiar arc-like feature with a length of $\sim$2000\,au, which was originally discovered by the HCO$^+$\,(3--2) observations \citep{Tokuda14}. Hydrodynamical simulation by \cite{Matsumoto15} suggested that gravitational torque from multiple objects promotes such a complex arc-like gas with a size scale of $\sim$1000\,au. \cite{Kuffmeier19} suggested that similar bridge-like structures are a transient phenomenon and considered to be a possible piece of evidence for multiple star formation. In summary, our frequency setting is reasonable to trace a density range of 10$^2$--10$^6$\,cm$^{-3}$ continuously and has the potential to investigate complex gas dynamics related to multiple star formation. 

\subsection{Observations and Data Reductions/Qualities}\label{obs_reduct}
Detailed data reduction processes and qualities in PROJ4 and PROJ-MC27 were presented in \cite{Tokuda19} and \cite{Tokuda18}, respectively. We summarize the observation settings and data qualities in Table\,\ref{tab:obsset} and describe PROJ6 as follows. We performed the ACA (the 7\,m array and the TP (Total Power) array) observations toward the Taurus dense cores (Sect.\,\ref{obs:design}) between 2018 October and 2019 March. The covered $uv$ range of the 7\,m array was 5.4--29\,k$\lambda$. There were three pointings per target with the same pattern (see also Figure\,\ref{fig:all_core}). The integration time (on-source time) per pointing was $\sim$8.5\,minutes. The bandpass and flux calibrator was the quasar J0423--0120. We observed J0510+1800 or J0336+3218 as phase calibrators once every $\sim$8\,minutes during the observations. There were four spectral windows for molecular line observations with a bandwidth of 62.5\,MHz and a spectral resolution of 61.0\,kHz, targeting N$_2$D$^{+}$, $^{12}$CO, $^{13}$CO, and C$^{18}$O (Sect.\,\ref{obs:freq}). We used two spectral windows, whose bandwidth and resolution were 2000\,MHz and 15.6\,MHz, respectively, to obtain continuum images. 
The central frequencies of the two bands were 218.0 and 232.5\,GHz. We individually processed the two bands of two targets, MC1 and MC35, following the procedure described in the next paragraph, as a preliminary analysis. After we confirmed that the two bands reproduce almost the same result (see Sect.\,\ref{App:Two_cont} in Appendix\,\ref{AppA}), we combined the two bands to enhance the image sensitivity. The effective frequency of the final processed continuum images is 225.3\,GHz ($\sim$1.3\,mm).

We performed the data reduction using the Common Astronomy Software Application (CASA) package \citep{McMullin07} version 5.6.0. In the imaging process, we used the \texttt{tclean} task with the \texttt{multi-scale} deconvolver \citep{Kepley20} to recover the extended emission. The imaging grid was set to have square pixels of 1\farcs0 width, and the scales of the \texttt{multi-scale} clean are 0, 6, and 18 pixels. The weighting scheme was $``$Natural.$"$ We manually selected the emission mask (clean box) and continued the deconvolution process until the intensity of the residual image reached the $\sim$1$\sigma$ noise level. We did not apply the self-calibration process, because the continuum emission of the prestellar sources is quite weak. Note that we applied the same method and parameters to the other two different data sets (PROJ4 and PROJ-MC27). Three targets (MC5N, MC31, and MC33bS) were observed in both PROJ4 and PROJ6 and had confirmed the continuum detection in the two bands, but we used the PROJ6 (1.3\,mm) data alone throughout the analyses in this manuscript. We manually selected emission-free pixels from each continuum image to estimate their rms noise levels. Table\,\ref{tab:obsset} summarizes the resultant beam sizes and sensitivities (see also the individual properties of each target in Table\,\ref{table:beam_rms} in Appendix\,\ref{AppA}).

In the present paper, we normally did not use the TP molecular line data to focus on compact features revealed by the 7\,m array continuum observation alone and to keep the spatial frequency range of both the line and continuum data similar. Only in Sect.\,\ref{dis15sub}, we used the combined 7\,m and TP array images obtained by the \texttt{feather} task in CASA to further obtain evidence for the realistic substructures within two prestellar cores, MC1 and MC7.

\begin{table}[htpb]
\begin{flushleft}
\caption{Observation settings and qualities}
\label{tab:obsset}
\begin{tabular}{lllllll}
\hline
Project name & Code & Continuum & Molecular lines & Beam size                                                   & Continuum rms & Line rms$^{\rm b}$ \\ \hline
PROJ4$^{\rm a}$        &  2016.1.00928.S  & 1.2\,mm         & H$^{13}$CO$^{+}$, H$^{13}$CN\,(3--2)  & 7\farcs4\,$\times$\,5\farcs2 & $\sim$0.6\,mJy\,beam$^{-1}$ & $\sim$0.07\,K \\
PROJ6$^{\rm a}$        &  2018.1.00756.S  & 1.3\,mm         & $^{12}$CO, $^{13}$CO, C$^{18}$O\,(2--1), N$_2$D$^{+}$\,(3--2) & 6\farcs8\,$\times$\,6\farcs3 & $\sim$0.4\,mJy\,beam$^{-1}$ & $\sim$0.05\,K \\
PROJ-MC27    &  2015.1.00340.S  & 1.3\,mm         & $^{13}$CO, C$^{18}$O\,(2--1), N$_2$D$^{+}$\,(3--2) & 7\farcs2\,$\times$\,5\farcs8  & $\sim$0.3\,mJy\,beam$^{-1}$ & $\sim$0.03\,K \\ \hline
\end{tabular}
\end{flushleft}
\tablenotetext{\rm a}{The beam size and rms are the average values among the observed targets. Table\,\ref{table:beam_rms} in Appendix gives the individual ones.}
\tablenotetext{\rm b}{The velocity resolution is $\sim$0.1\,km\,s$^{-1}$.}
\end{table}

\section{Results} \label{sec:results}
\begin{figure}[htbp]
\includegraphics[width=180mm]{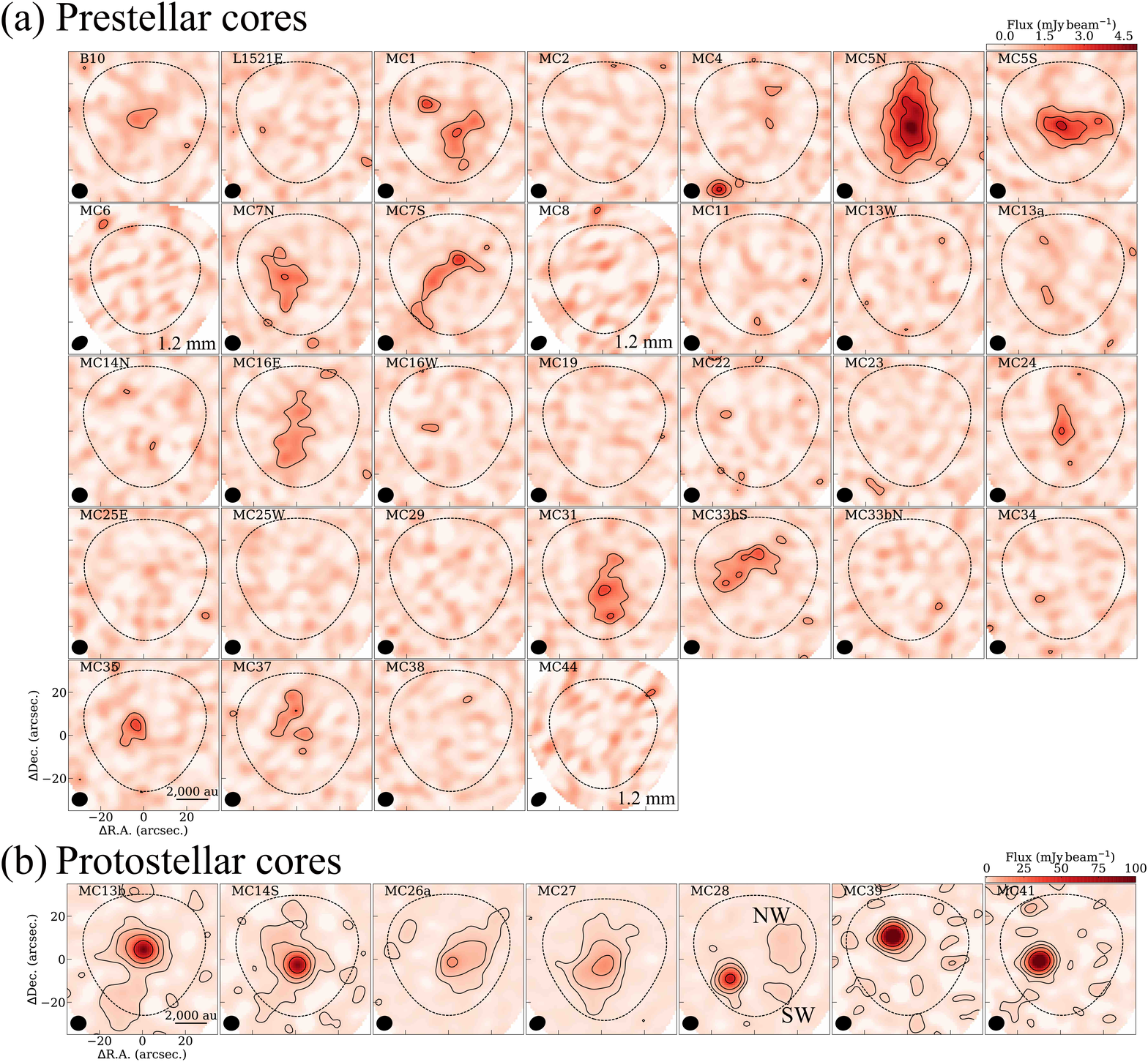}
\caption{1.3\,mm and 1.2\,mm continuum distributions of prestellar/protostellar cores in Taurus. (a) Continuum maps of the prestellar sources. For MC6, MC8, and MC44, the observed wavelength is 1.2\,mm, and we indicated it at the lower right corners. Table\,\ref{tab:target} gives the central coordinates of each target. Dashed black lines indicate where the mosaic sensitivity falls to 50\%. The black contours show the continuum emission with the contour levels of [3$\sigma$, 6$\sigma$, 9$\sigma$]. The ellipses in the lower left corner in each panel give the angular resolutions. Noise levels and beam sizes of each map are shown in Table\,\ref{table:beam_rms}. Note that the primary beam attenuation is not corrected for the display purpose. (b) Same as (a) but for the protostellar sources. The contour levels are [3$\sigma$, 10$\sigma$, 30$\sigma$, 100$\sigma$].
\label{fig:all_core}}
\end{figure}

Figure \ref{fig:all_core} shows 1.2/1.3\,mm continuum distributions at the innermost part of prestellar/protostellar cores in Taurus. In this section, we describe and characterize the properties of these objects based on the continuum data.

\subsection{Prestellar Cores} \label{results:prestellar}
Figure \ref{fig:all_core} shows the 7\,m array observations in 1.2/1.3\,mm continuum toward all of the observed sources. We could not detect significant ($>$3$\sigma$) emission in 20 prestellar cores (MC2, MC4, MC6, MC8, MC11, MC13a, MC13W, MC14N, MC16W, MC19, MC22, MC23, MC25E, MC25W, MC29, MC33bN, MC34, MC38, MC44, and L1521E). 
On the contrary, we confirmed continuum detection toward 12 prestellar cores (MC1,MC5N, MC5S, MC7N, MC7S, MC16E, MC24, MC31, MC33bS, MC35, MC37, and B10). The detection rate is about one-third of the observed prestellar cores, which is quite high compared to the previous unbiased surveys with a successful detection of zero or one sources toward Chamaeleon\,I and $\rho$ Ophiuchus using the ALMA 12\,m array alone \citep{Dunham16,Kirk17}, because the larger beam and recovering scales in our observations enable us to be more sensitive to the lower-density regions of the cores (see also discussions in Sect.\,\ref{dis2}). The peak ($F_{\rm peak}$) and total continuum flux ($F_{\rm \nu}$) within the observed field are 2--5\,mJy\,beam$^{-1}$ and 5--63\,mJy, respectively, as summarized in Table\,\ref{table:1mm_pre}. We performed 2D Gaussian fittings to the observed continuum distributions in the image plane after the primary beam correction to derive the central coordinate, major/minor axis (FWHM), and position angle.
Note that, in the observed field, there is a large amount of missing flux, more than 80\%, with respect to the total flux obtained by single-dish telescopes (see the MC5N case in \citealt{Tokuda19}).

We derived the mass of the central regions traced by the 7\,m array continuum observations using the following equation, 
\begin{equation}\label{eq:Mass}
 M = \frac{D^2F_{\rm 1.3\,mm}}{\kappa_{\rm 1.3\,mm}B(T_{\rm d})} 
\end{equation}
where $D$ is the distance to the sources, $\kappa_{\rm 1.3\,mm}$ is the dust opacity at 1.3\,mm, and $B$($T_{\rm d}$) is the Planck function at a dust temperature $T_{\rm d}$. We applied recent distance measurements depending on the region \citep{Galli18} and $\kappa_{\rm 1.3\,mm}$ = 0.005\,cm${^2}$\,g$^{-1}$ for prestellar cores \citep{Ossenkopf94,Preibisch93}. We used gas kinematic temperatures derived from the early NH$_{3}$ observations (see Table\,\ref{tab:target}), as $T_{\rm d}$, assuming that the temperatures of gas and dust are well coupled in the dense regions and a uniform temperature distribution inside the 7\,m array observation fields. Table\,\ref{table:1mm_pre} gives the derived mass. The uncertainty of the mass estimation mainly arises from the assumption of $\kappa_{\rm 1.3\,mm}$ due to the grain growth at the central regions of the cores. However, \cite{Bracco17} reported that the dust emissivity index, $\beta$, does not change too much across core radii down to 1000\,au at the prestellar cores in the Taurus B213 filament. Although \cite{Chac17} suggested that the opacity can increase by a factor of $\sim$2, which is likely traced by ALMA's resolution, such effect is limited to a small radius, a few hundred astronomical units. The grain growth thus does not strongly affect the mass estimation with the 7\,m array beam size.

The estimated masses ($M_{\rm obs}$) are $\sim$10$^{-2}$--10$^{-1}$\,$M_{\odot}$ (see Table\,\ref{table:1mm_pre}), indicating that the 7\,m array observations likely trace compact/dense materials at the center of the cores, but show only small fractions of the entire cores themselves whose total masses are $\sim$1--10\,$M_{\odot}$. We estimated the average H$_2$ number densities using the following equation: $n(\rm H_2)$~=~3$M_{\rm obs}$/4$\pi\mu$m$_{\rm H}R_{\rm obs}^{3}$, where $\mu$ is the molecular weight per hydrogen (2.8), $m_{\rm H}$ is the H atom mass, and $R_{\rm obs}$ is the observed radius (=$\sqrt{\rm Major*Minor}$/2) adopting each distance (Table\,\ref{tab:target}). The derived densities are $\sim$(3--10)$\times$10$^5$\,cm$^{-3}$ (see Table\,\ref{table:1mm_pre}). The identification of the high-density part in the molecular clouds based on millimeter continuum emission alone is sometimes risky, because there may be possible contaminations from unrelated objects, such as distant galaxies \citep[e.g.,][]{Tamura15,Tokuda16}. Examining the presence or absence of detection of high-density gas tracers allows us to confirm whether the continuum emission is indeed arising from the deeply embedded parts in dense cores. We confirmed C$^{18}$O emission toward all of the PROJ6 targets, and most of the continuum peaks have the N$_2$D$^+$ emission at more than the 3$\sigma$ level (see Table\,\ref{table:1mm_pre}). We thus conclude that the observed 1.3\,mm continuum is arising from the cold high-density parts of the cores. Note that we found an additional source with a flux of 19.3\,mJy at the edge of the observed field in MC4. The position corresponds to a known YSO candidate, J041412.29+280837.0 \citep{Gutermuth09,Rebull11}.

The spatial distributions of the 1.3\,mm continuum are diverse: some cores show single peaks while the others contain multiple local peaks. For the first time, we have revealed such a complex substructure toward a large fraction of the sample in Taurus prestellar cores. However, \cite{Caselli19} cautioned that the interferometric artifacts cause fake substructures even if we observe a smoothed distribution. We discuss this possibility in Sect.\,\ref{dis1} further. 

\begin{table}[htbp]
\caption{1.3 mm continuum properties of prestellar cores}
\label{table:1mm_pre}
\begin{center}
\footnotesize
\begin{tabular}{lccccccccccc} \hline
Name  & RA         & Dec       & Major     & Minor     & P.A.   & $F_{\rm peak}$     & $F_\nu$ & Mass          & $n$(H$_2$)  & Category &  N$_2$D$^+$ \\ 
      & (J2000.0)  & (J2000.0) & (arcsec.) & (arcsec.) & (deg.) & (mJy\,beam$^{-1}$) & (mJy)   & ($\times$10$^{-2}$\,$M_{\odot}$) & ($\times$10$^{5}$\,cm$^{-3}$) &          &    \\ \hline
MC1   & 04$^{\rm h}$04$^{\rm m}$47\fs7 & +26\arcdeg09\arcmin15\farcs8 & 32.4  & 16.8  & 161.8 & 3.3 & 20.5 & 3.5 & 3.3 &II & Y  \\
MC2   & $\cdots$  & $\cdots$  & $\cdots$ & $\cdots$ & $\cdots$ & $\cdots$   & $\cdots$  & $\cdots$   & $\cdots$ &I & N  \\
MC4   & $\cdots$  & $\cdots$  & $\cdots$ & $\cdots$ & $\cdots$ & $\cdots$   & $\cdots$  & $\cdots$   & $\cdots$ &I & N  \\
MC5N  & 04$^{\rm h}$17$^{\rm m}$42\fs0 & +28\arcdeg08\arcmin44\farcs7 & 38.9  & 17.5  & 173.1 & 5.4 & 63.4  & 12.8 & 8.1 &III & Y  \\
MC5S  & 04$^{\rm h}$17$^{\rm m}$42\fs6 & +28\arcdeg06\arcmin00\farcs2 & 31.0  & 13.0  & 88.9  & 3.9 & 26.7  & 5.0  & 7.0 &III & Y  \\
MC6   & $\cdots$  & $\cdots$  & $\cdots$ & $\cdots$ & $\cdots$ & $\cdots$   & $\cdots$  & $\cdots$   & $\cdots$ &I & $\cdots$ \\
MC7N  & 04$^{\rm h}$18$^{\rm m}$04\fs1 & +28\arcdeg22\arcmin56\farcs7 & 21.8  & 14.8  & 18.0  & 2.5 & 14.0  & 2.6 & 5.1 &II & Y  \\
MC7S  & 04$^{\rm h}$18$^{\rm m}$03\fs0 & +28\arcdeg22\arcmin21\farcs0 & 35.7  & 8.6   & 134.2 & 3.2 & 21.2 & 4.0 & 8.4 &III & Y  \\
MC8   & $\cdots$  & $\cdots$  & $\cdots$ & $\cdots$ & $\cdots$ & $\cdots$   & $\cdots$  & $\cdots$   & $\cdots$ &I & $\cdots$ \\
MC11  & $\cdots$  & $\cdots$  & $\cdots$ & $\cdots$ & $\cdots$ & $\cdots$   & $\cdots$  & $\cdots$   & $\cdots$ &I & Y  \\
B10   & 04$^{\rm h}$17$^{\rm m}$50\fs2 & +27\arcdeg56\arcmin10\farcs7 & 22.5  & 10.7  & 24.5  & 2.0 & 5.2  & 1.1 & 3.1 &II & Y  \\
MC13W & $\cdots$  & $\cdots$  & $\cdots$ & $\cdots$ & $\cdots$ & $\cdots$   & $\cdots$  & $\cdots$   & $\cdots$ &I & N  \\
MC13a & $\cdots$  & $\cdots$  & $\cdots$ & $\cdots$ & $\cdots$ & $\cdots$   & $\cdots$  & $\cdots$   & $\cdots$ &I & N \\
MC14N & $\cdots$  & $\cdots$  & $\cdots$ & $\cdots$ & $\cdots$ & $\cdots$   & $\cdots$  & $\cdots$   & $\cdots$ &I & N  \\
MC16E & 04$^{\rm h}$21$^{\rm m}$21\fs2 & +26\arcdeg59\arcmin43\farcs5 & 36.7  & 13.2  & 165.7 & 2.5 & 17.8  & 5.1 & 3.0 & II & Y  \\
MC16W & $\cdots$  & $\cdots$  & $\cdots$ & $\cdots$ & $\cdots$ & $\cdots$   & $\cdots$  & $\cdots$   & $\cdots$ &I & N  \\
MC19  & $\cdots$  & $\cdots$  & $\cdots$ & $\cdots$ & $\cdots$ & $\cdots$   & $\cdots$  & $\cdots$   & $\cdots$ &I & N  \\
MC22  & $\cdots$  & $\cdots$  & $\cdots$ & $\cdots$ & $\cdots$ & $\cdots$   & $\cdots$  & $\cdots$   & $\cdots$ &I & N  \\
MC23  & $\cdots$  & $\cdots$  & $\cdots$ & $\cdots$ & $\cdots$ & $\cdots$   & $\cdots$  & $\cdots$   & $\cdots$ &I & N  \\
MC24  & 04$^{\rm h}$26$^{\rm m}$35\fs4 & +24\arcdeg41\arcmin44\farcs4 & 26.3  & 8.8   & 9.2   & 2.7 & 12.5   & 3.0 & 5.2 &II & N  \\
MC25E & $\cdots$  & $\cdots$  & $\cdots$ & $\cdots$ & $\cdots$ & $\cdots$   & $\cdots$  & $\cdots$   & $\cdots$ &I & Y  \\
MC25W & $\cdots$  & $\cdots$  & $\cdots$ & $\cdots$ & $\cdots$ & $\cdots$   & $\cdots$  & $\cdots$   & $\cdots$ &I & N \\
MC28NW& 04$^{\rm h}$29$^{\rm m}$22\fs8 & +24\arcdeg33\arcmin15\farcs0 & 21.8  & 4.3   & 2.7   & 4.3 & 23.1  & 3.4 & 44.6 & $\cdots^{\rm a}$ & N  \\
MC28SW& 04$^{\rm h}$29$^{\rm m}$22\fs0 & +24\arcdeg33\arcmin53\farcs5 & 11.6  & 10.5  & 58.0  & 4.6 & 10.4  & 1.5 & 13.5 & $\cdots^{\rm a}$ & N  \\
MC29  & $\cdots$  & $\cdots$  & $\cdots$ & $\cdots$ & $\cdots$ & $\cdots$   & $\cdots$ & & & I & N  \\
MC31  & 04$^{\rm h}$31$^{\rm m}$55\fs3 & +24\arcdeg32\arcmin48\farcs4 & 25.8  & 14.6  & 4.3   & 3.5 & 24.3  & 5.4 & 8.8 & III & Y  \\
MC33bS& 04$^{\rm h}$32$^{\rm m}$43\fs6 & +24\arcdeg23\arcmin15\farcs1 & 39.5  & 12.6  & 123.0 & 2.8 & 16.8  & 3.3 & 3.6 & II & N  \\
MC33bN& $\cdots$  & $\cdots$  & $\cdots$ & $\cdots$ & $\cdots$ & $\cdots$   & $\cdots$ & $\cdots$   & $\cdots$ & I  & Y  \\
MC34  & $\cdots$  & $\cdots$  & $\cdots$ & $\cdots$ & $\cdots$ & $\cdots$   & $\cdots$ & $\cdots$   & $\cdots$ & I  & N  \\
MC35  & 04$^{\rm h}$35$^{\rm m}$37\fs9 & +24\arcdeg09\arcmin20\farcs0 & 12.2  & 10.1  & 169.0 & 2.8 & 7.8 & 1.3  & 11.6 & $\cdots^{\rm b}$ &  Y  \\
MC37   & 04$^{\rm h}$39$^{\rm m}$18\fs0 & +25\arcdeg52\arcmin34\farcs1 & 23.9  & 9.2   & 153.7 & 2.4 & 9.2 & 1.6 & 4.4 & II   & N  \\
MC38   & $\cdots$  & $\cdots$  & $\cdots$ & $\cdots$ & $\cdots$ & $\cdots$   & $\cdots$ & $\cdots$   & $\cdots$ & I   & N  \\
MC44   & $\cdots$  & $\cdots$  & $\cdots$ & $\cdots$ & $\cdots$ & $\cdots$   & $\cdots$ & $\cdots$   & $\cdots$ & I  & $\cdots$ \\
L1521E & $\cdots$  & $\cdots$  & $\cdots$ & $\cdots$ & $\cdots$ & $\cdots$   & $\cdots$ & $\cdots$   & $\cdots$ & I   & N \\ \hline
\end{tabular}
\end{center}
\tablenotetext{\rm a}{We did not categorize the starless peaks in the MC28 protostellar core.}
\tablenotetext{\rm b}{A FHSC candidate (see Paper II).}
\end{table}

\subsection{Protostellar Cores} \label{results:protostellar}
We observed seven protostellar cores (MC13b, MC14S, MC26a, MC27, MC28, MC39, and MC41) with the 7\,m array. We detected 1.3\,mm continuum emission in all targets. The observed total and peak fluxes are 62--225\,mJy and $\sim$14--202\,mJy\,beam$^{-1}$, respectively (Table\,\ref{table:1mm_proto}), which are about an order of magnitude higher than those in the prestellar sources. Although the early single-dish observations also confirmed this trend \citep[e.g.,][]{Motte01}, the filtering-out effect, due to the interferometric observation, enhanced the intensity contrast between the two evolutionary stages. For example, the early study in MC14S/N provided just a factor of 3 difference in the peak 1.3\,mm intensities between the two sources \citep{Bracco17}. However, our new observations could not detect continuum emission in the prestellar source. We compiled higher-resolution interferometric observations at the same frequency from the literature for all protostellar sources (Table\,\ref{table:1mm_proto}) to determine the disk contamination. The much longer baseline observations enable us to reveal further compact emission, which is mostly arising from the protostellar disk. Although the fraction of protostellar disk contribution ($F_{\rm disk}$) to the total flux covered by the 7\,m array depends from source to source, most of the flux is dominated by $F_{\rm disk}$ with a range of 35--90\% except for MC27. 

MC27 is a unique source in our sample containing a class\,0 VeLLO \citep{Bourke06,Terebey09} in the L1521 region (Figure\,\ref{fig:Taurusmap}). Note that there is another VeLLO, IRAM 04191--1522 \citep[see][]{Andre99,Dunham06}, which is located at the outside of the Taurus main cloud. Among the observed protostellar cores, MC27 shows the weakest peak flux, and the contribution from the protostellar disk is $\sim$1\,mJy \citep[see also the 12\,m array observations in][]{Tokuda14,Tokuda16}. 

In MC28, we found two starless peaks (MC28NW/SW) within the $>$50\% sensitivity field, and their locations are away from the class I binary source, IRAS\,04263+2426 \citep{Chandler98,Roccatagliata11}. The continuum peaks have the C$^{18}$O emission with a central velocity of $\sim$6\,km\,s$^{-1}$, which is similar to that of the binary source, indicating that these sources belong to the same system.

\begin{table}[htbp]
\caption{1.3 mm continuum properties of protostellar cores}
\label{table:1mm_proto}
\begin{center}
\footnotesize
\begin{tabular}{lccccccccccc}\hline
Name  & Infrared Source$^{\rm a}$ & R.A.  & Decl. & Major  & Minor   & P.A.   & $F_{\rm peak}$ & $F_{\nu}$ & $F_{\rm disk}$ & N$_2$D$^{+}$  \\
   &   & (J2000.0) & (J2000.0) & (arcsec.) & (arcsec.) & (deg.) & (mJy\,beam$^{-1}$) & (mJy) & (mJy) &  \\ \hline 
MC13b & IRAS\,04166+2706 & 04$^{\rm h}$19$^{\rm m}$42\fs5 & +27\arcdeg13\arcmin36\farcs1 & 8.8  & 7.8  & 78.0  & 97  & 187 & 66$^{\rm b}$ & Y  \\
MC14S & IRAS\,04169+2702 & 04$^{\rm h}$19$^{\rm m}$58\fs5 & +27\arcdeg09\arcmin56\farcs8 & 7.6  & 7.4  & 72.0  & 92  & 147 & 88$^{\rm c}$  & Y  \\
MC26a & IRAS\,04248+2612 & 04$^{\rm h}$27$^{\rm m}$57\fs0 & +26\arcdeg19\arcmin18\farcs4 & 21.5 & 12.7 & 122.1 & 14   & 62  & $\cdots$ & N  \\
MC27  & L1521F-IRS       & 04$^{\rm h}$28$^{\rm m}$39\fs1 & +26\arcdeg51\arcmin32\farcs8 & 19.4 & 13.5 & 143.4 & 15 & 78 & 1$^{\rm d}$  & Y  \\
MC28  & IRAS\,04263+2426 & 04$^{\rm h}$29$^{\rm m}$23\fs8 & +24\arcdeg33\arcmin00\farcs3 & 7.3  & 6.8  & 121.4 & 76   & 93  & 84$^{\rm e}$ & N  \\
MC39  & IRAS\,04365+2535 & 04$^{\rm h}$39$^{\rm m}$35\fs2 & +25\arcdeg41\arcmin44\farcs3 & 7.2  & 6.4  & 102.0 & 202  & 225 & 180$^{\rm f}$   & N  \\
MC41  & IRAS\,04369+2539 & 04$^{\rm h}$39$^{\rm m}$55\fs8 & +25\arcdeg45\arcmin01\farcs5 & 7.1  & 6.3  & 109.1 & 179  & 201 & $\cdots$ & N \\ \hline
\end{tabular}
\end{center}
\tablenotetext{\rm a}{\cite{Kenyon90} for IRAS sources, \cite{Bourke06} for L1521F-IRS.}
\tablenotetext{\rm b}{Combined Array for Research in Millimeter-wave Astronomy (CARMA) observations with angualr resolution of 1\farcs03$\times$0\farcs90 by \cite{Eisner12}.}
\tablenotetext{\rm c}{Submillimeter Array (SMA) observations with an angular resolution of 1\farcs76$\times$1\farcs50 by \cite{Takakuwa18}.}
\tablenotetext{\rm d}{ALMA observations with angular resolution of 1\farcs3$\times$0\farcs8 by \cite{Tokuda16}.}
\tablenotetext{\rm e}{CARMA observations with an angular resolution of 1$\arcsec$ by \cite{Sheehan14}. Note that we combined individual fluxes from the binary source ($\sim$44\,mJy for source N and $\sim$40\,mJy for source S).}
\tablenotetext{\rm f}{ALMA observations with an angular resolution of 1\farcs01$\times$0\farcs87 by \cite{Aso15}.}
\end{table}

\section{Discussions} \label{sec:dis}

\subsection{Interpretations of Continuum Detection in Prestellar cores} \label{dis1}
Revealing the fragmentation/coalescence process of molecular cloud cores just before the onset of star formation is critical toward understanding the origin of binary/multiple star formation and determining the final stellar mass.  
For example, turbulent perturbations \citep[e.g.,][]{Offner10} may create a few hundred astronomical unit scale overdense regions locally. However, some of the magnetohydrodynamic simulations \citep{Matsumoto11} suggest that the gravitational collapse smoothed out the complex substructures that originated from turbulence. In this case, observations should show that prestellar cores do not have complex substructures (i.e., fragments), possibly leading to multiple objects. The exploration of such a spatial scale in low-mass dense cores is still incomplete (see Sect.\,\ref{intro:array}). The present ACA observations enable us to obtain crucial hints to understanding the mechanism of fragmentation and the evolution of the prestellar collapse phase. 

The 7\,m array measurements obtained indications of multiple local peaks with an intensity of a few mJy\,beam$^{-1}$ toward MC1, MC7N, MC7S, MC16E, MC31, MC33bS, and MC37. The typical separation among the peaks within a core is similar to the present beam size of $\sim$900\,au. The size scale is much smaller than the Jeans length with a density of $\sim$10$^6$\,cm$^{-3}$. If the substructures are real features, the Jeans instability is unlikely to form such fragments. However, we need careful treatments to interpret the substructures obtained by interferometers. \cite{Caselli19} cautioned that the incomplete cancellation of Fourier components could produce artificial substructures with a low-level contrast among the local peaks even if the real core has a constant-density profile at the center.

We evaluate the spatial distributions of what the 7\,m array observations are looking at the core centers and how realistic the multiple peaks are, by comparing the synthetic observations using CASA. We generated smoothed core models with a Plummer-like function as input models for the simulated observations. The formula of column density profile as a function of $r$, distance from the core center, is as follows:
\begin{equation}\label{eq:plummer}
 N_{\rm H_2}(r) = \frac{N_{\rm peak}}{(1+(r/R_{\rm flat})^2)^\frac{p-1}{2}} 
\end{equation}
There are three free parameters: peak H$_2$ column density ($N_{\rm peak}$), flattening radius ($R_{\rm flat}$), and asymptotic power index ($p$).
Based on the Herschel/SPIRE measurements (e.g., \citealt{Marsh16}) toward the 1.3\,mm continuum-detected objects with the 7\,m array (see Sect.\,\ref{sec:dis}), we adapted a fixed $p$ of 2 and measured $N_{\rm H_2}$ at a radius of $r$ = 6000\,au to determine $N_{\rm peak}$ using the equation\,(\ref{eq:plummer}). We prepared three sequences with different $N_{\rm H_2}$ ($r$ = 6000\,au), maximum = 3\,$\times$10$^{22}$\,cm$^{-2}$, average = 1\,$\times$10$^{22}$\,cm$^{-2}$, and minimum = 7\,$\times$10$^{21}$\,cm$^{-2}$ to mimic the observed sample. For each column density set, the input $R_{\rm flat}$ are 8000\,au, 2500\,au, 1500\,au, 900\,au, and 300\,au, which roughly corresponds to the Jeans length (see the definition in Sect\,\ref{intro:array}) of 10$^{5}$\,cm$^{-3}$, 10$^{6}$cm$^{-3}$, 3\,$\times$\,10$^{6}$\,cm$^{-3}$, 10$^{7}$\,cm$^{-3}$, and 10$^{8}$\,cm$^{-3}$ gas, respectively. 
Note that we assumed $\kappa_{\rm 1.3\,mm}$ = 0.005\,cm$^{2}$\,g$^{-1}$ and $T_{\rm d}$ = 10\,K to convert H$_2$ column densities to 1.3\,mm continuum fluxes (see also the justifications in Sect.\,\ref{results:prestellar}). In order to investigate the geometric effect of the simulated profiles, we considered parametric models with aspect ratios (see Equation (5) in \citealt{Caselli19}). We adapted two aspect ratios of 1.0 and 1.8. The former is called as the fixed model. The latter is a parametric model, which mimics the observed aspect ratio of MC31.
In the CASA simulator, we used the \texttt{simobserve} task to generate simulated visibilities at a central frequency of 225\,GHz and a similar integration time of our real observations (see Sect.\,\ref{sec:obs}) with the same 7\,m array configuration in Cycle\,6. The central sky coordinate was the same as that of MC27, which is close to the central position of the Taurus main cloud (see Figure\,\ref{fig:Taurusmap}). We made the simulated images using the \texttt{tclean} task with the same parameters that we applied to the real observations (see Sect.\,\ref{obs_reduct}).

\begin{figure}[htbp]
\includegraphics[width=180mm]{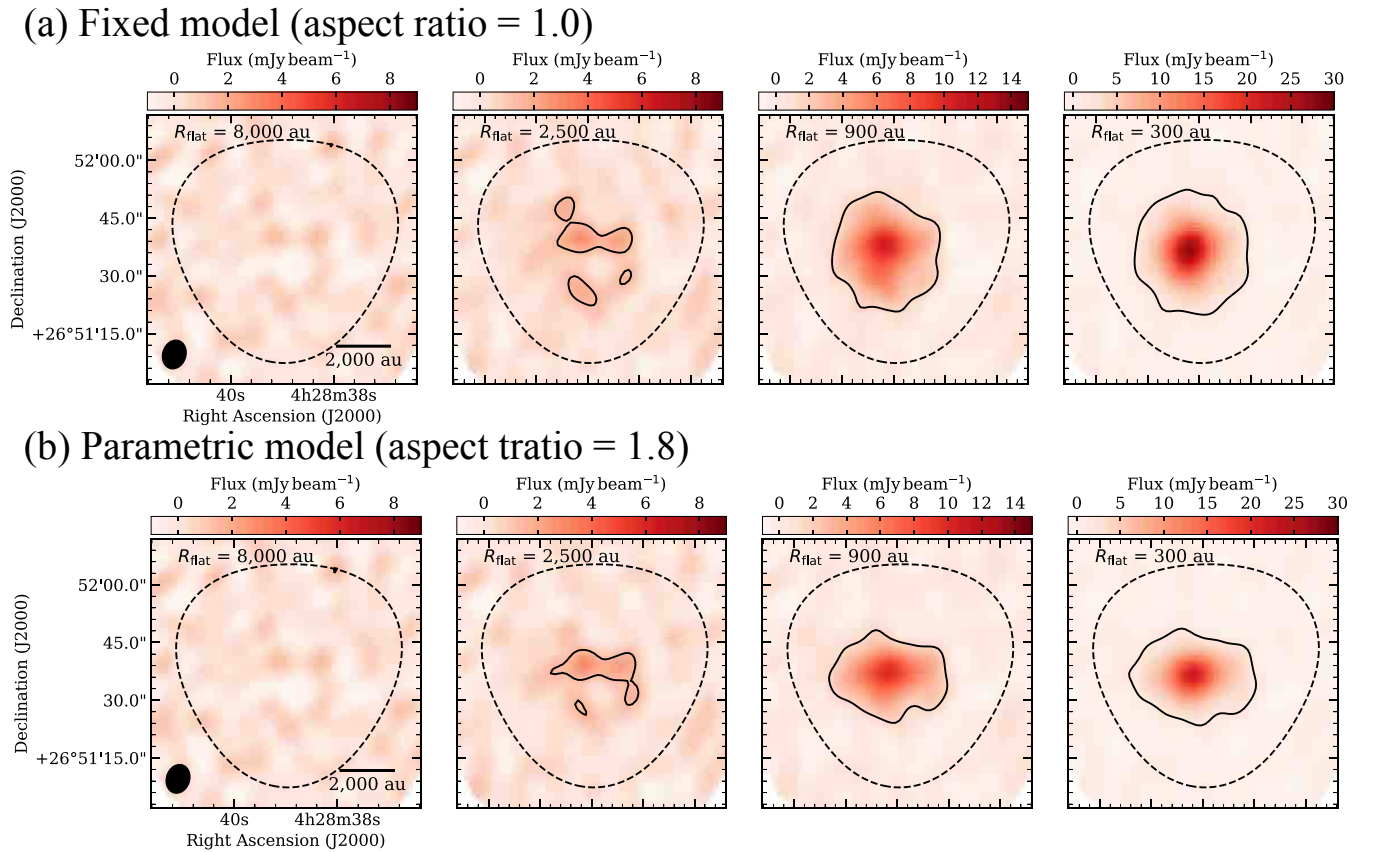}
\caption{Synthetic 7\,m array observations in 1.3\,mm continuum toward smoothed cores (see the text) with different flattening radii ($R_{\rm flat}$), indicated in the upper-left corner in each panel. The angular resolution, 7\farcs6\,$\times$\,6\farcs0 is given by the black ellipse in the lower-left corner in the leftmost panels. The dashed lines indicate where the mosaic sensitivity falls to 50\%. The contour level in each panel is the 3$\sigma$ noise level, where 1$\sigma$ is 0.49\,mJy\,beam$^{-1}$.
\label{fig:sim_obs}}
\end{figure}

\begin{figure}[htbp]
\begin{center}
\includegraphics[width=160mm]{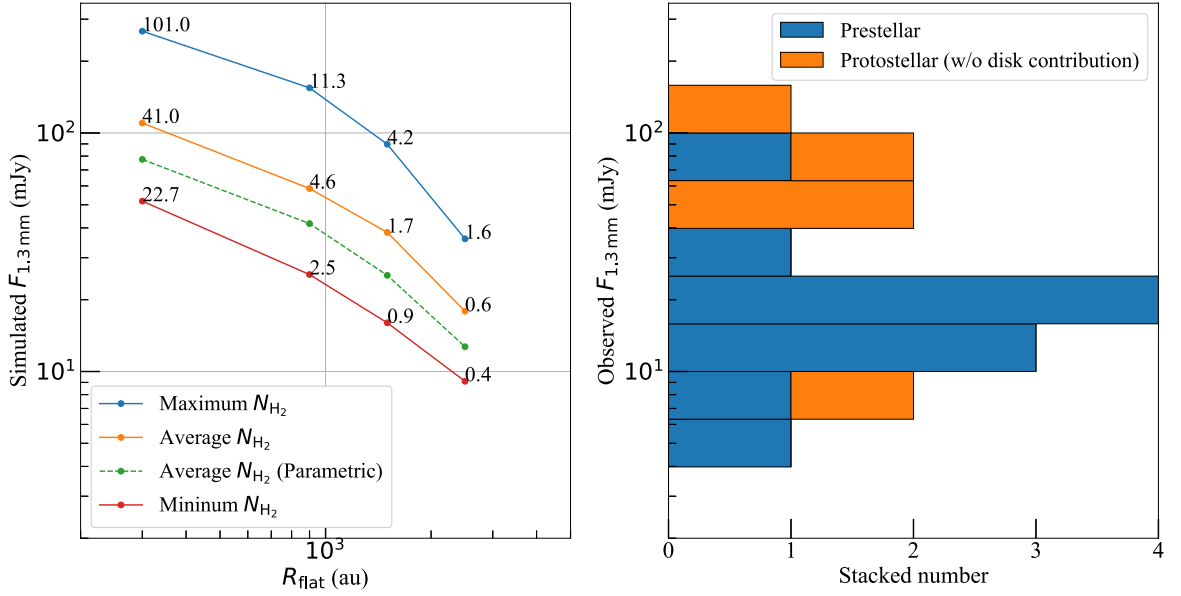}
\end{center}
\caption{(a) Synthetic 1.3\,mm continuum flux of simulated cores with Plummer-like functions. The absolute flux depends on the column densities, flattening radius, and aspect ratio of simulated cores (see the text). Estimated central densities in units of $\times$\,10$^{6}$\,cm$^{-3}$ are denoted at the vicinities of each symbol. The green symbols show the results from parametric models with an aspect ratio of 1.8 (see the text). (b) Observed 1.3\,mm continuum flux obtained from the 7\,m array observations. Blue and orange histograms show the stacked number of sources detected in prestellar and protostellar cores. Note that we subtracted the contributions of protostellar disks from the total flux obtained by the 7\,m array (see Table\,\ref{table:1mm_proto}).    
\label{fig:Flux_plot}}
\end{figure}

Figure\,\ref{fig:sim_obs} shows the results of the 7\,m array synthetic observations for the cases of average H$_2$ column density in the fixed (aspect ratio = 1.0) and parametric (aspect ratio = 1.8) models. We could not detect any significant (above 3$\sigma$) emission at the $R_{\rm flat}$ of 8000\,au cases. For the other radii models, the fixed models are close to circularly symmetric shapes as a whole, while the parametric models have elongated distributions in the east--west direction. Although the integrated fluxes, which were measured using the same method we used for our real observations (see Sect.\,\ref{sec:results}), depend on the absolute column density of the input models as shown in Figure\,\ref{fig:Flux_plot} (a), the same $R_{\rm flat}$ models well reproduce similar spatial distributions to each other. We then compared the real and synthetic observations. The case of $R_{\rm flat}$ of 2500\,au shown in Figure\,\ref{fig:sim_obs} is very similar to the observed substructures, as shown in Figure\,\ref{fig:all_core}. This means that the 7\,m array measurements can artificially produce multiple peaks whose angular separation is similar to the observed beam size, even if we observe a smooth distribution inside the observed fields (see also in the Appendix \ref{Ap:Obs_model}). Therefore, we cannot merely claim that the observed substructure is a substantial piece of evidence of fragmentation of prestellar cores. We further discuss the fidelity of the internal substructures by considering the molecular line emission in Sect\,\ref{dis15sub}.

Figure\,\ref{fig:Flux_plot} (b) shows the histogram of the observed 1.3\,mm continuum flux at pre-/protostellar cores. We tentatively estimated the central volume density using the $N_{\rm flat}$/$R_{\rm flat}$ of the input models, where $N_{\rm flat}$ is the H$_2$ column density at $R_{\rm flat}$, as shown in each plot in Figure\,\ref{fig:Flux_plot} (a). Note that a similar method was also applied to real observations of centrally concentrated prestellar cores \citep[e.g.,][]{Ward-Thompson99}. The simulated 1.3\,mm flux of the minimum column density model with $R_{\rm flat}$ of 2500\,au gives us a detection limit of the central density, $\sim$4\,$\times$\,10$^{5}$\,cm$^{-3}$, which is consistent with our real observations. Only a few cores (MC5N, MC5S, MC7S, and MC31) have a relatively strong continuum flux more than $\sim$20\,mJy, and their central densities are likely more than 8\,$\times$\,10$^{5}$\,cm$^{-3}$ (see also Table\,\ref{table:1mm_pre}), which is higher than the other sources. The 1.3\,mm fluxes of the bright prestellar sources are comparable to those in protostellar cores without their disk contributions, as shown in Figure\,\ref{fig:Flux_plot} (b). This result can be an indirect piece of evidence that the bright cores are more evolved than the other weak cores. 

\subsection{Candidates with Internal Substructures in Prestellar Cores} \label{dis15sub}

\begin{figure}[htbp]
\includegraphics[width=180mm]{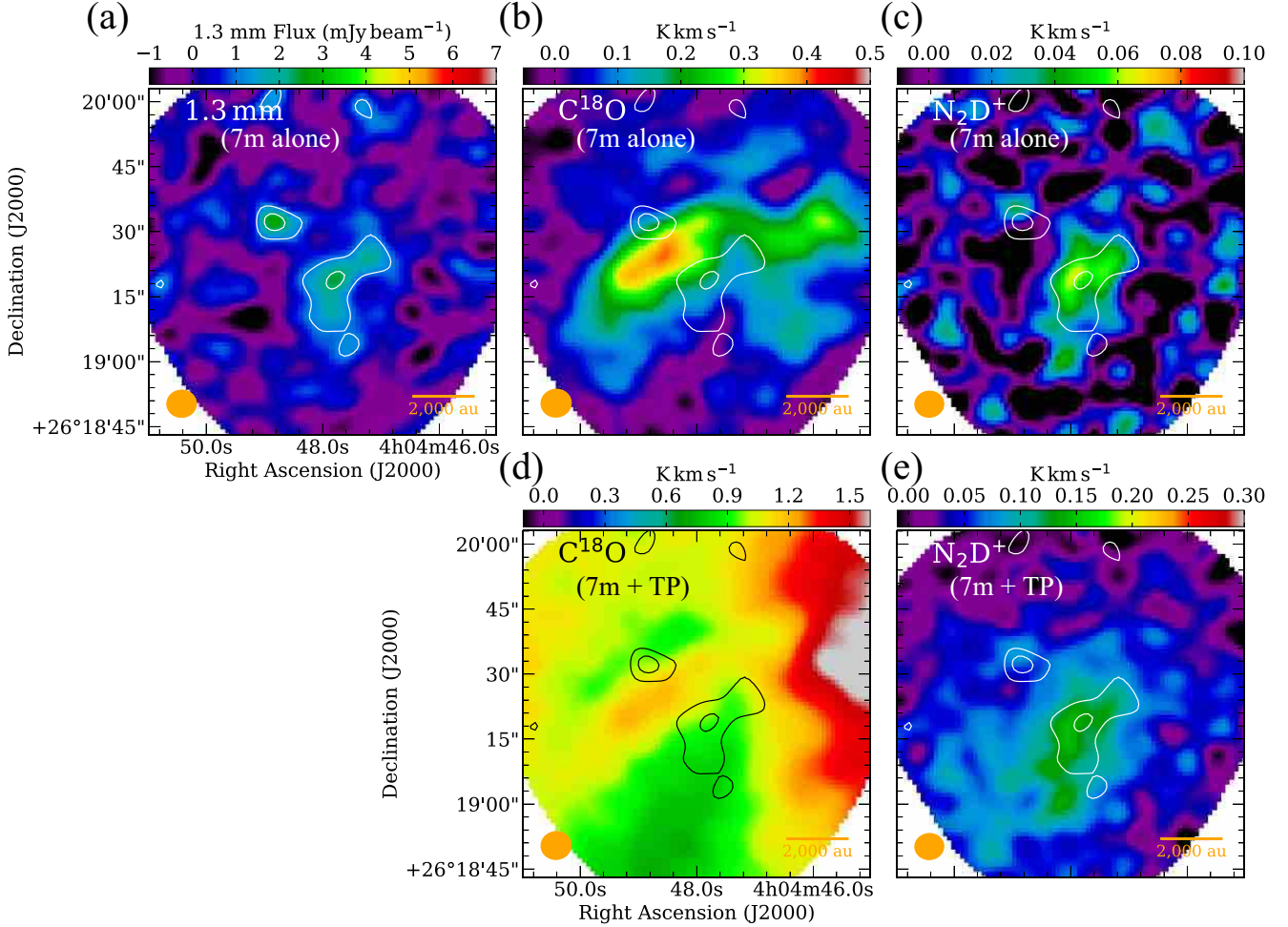}
\caption{The ACA (7\,m + TP array) observations in dust and molecular lines of MC1. (a) Color-scale image and white contours show the 1.3\,mm continuum emission obtained by the 7\,m array alone. The contour levels are the same as Figure\,\ref{fig:all_core}. The ellipse in each lower-left corner gives the angular resolutions. Color-scale images in panels (b) and (c) show the integrated intensity maps of C$^{18}$O\,(2--1) and N$_2$D$^+$\,(3--2), respectively, obtained by the 7\,m array alone. The contours and levels are the same as panel (a). (d) and (e) Same as panels (b) and (c) but for the combined image of the 7\,m + TP array. \label{fig:MC1_mol}}
\end{figure}

\begin{figure}[htbp]
\includegraphics[width=180mm]{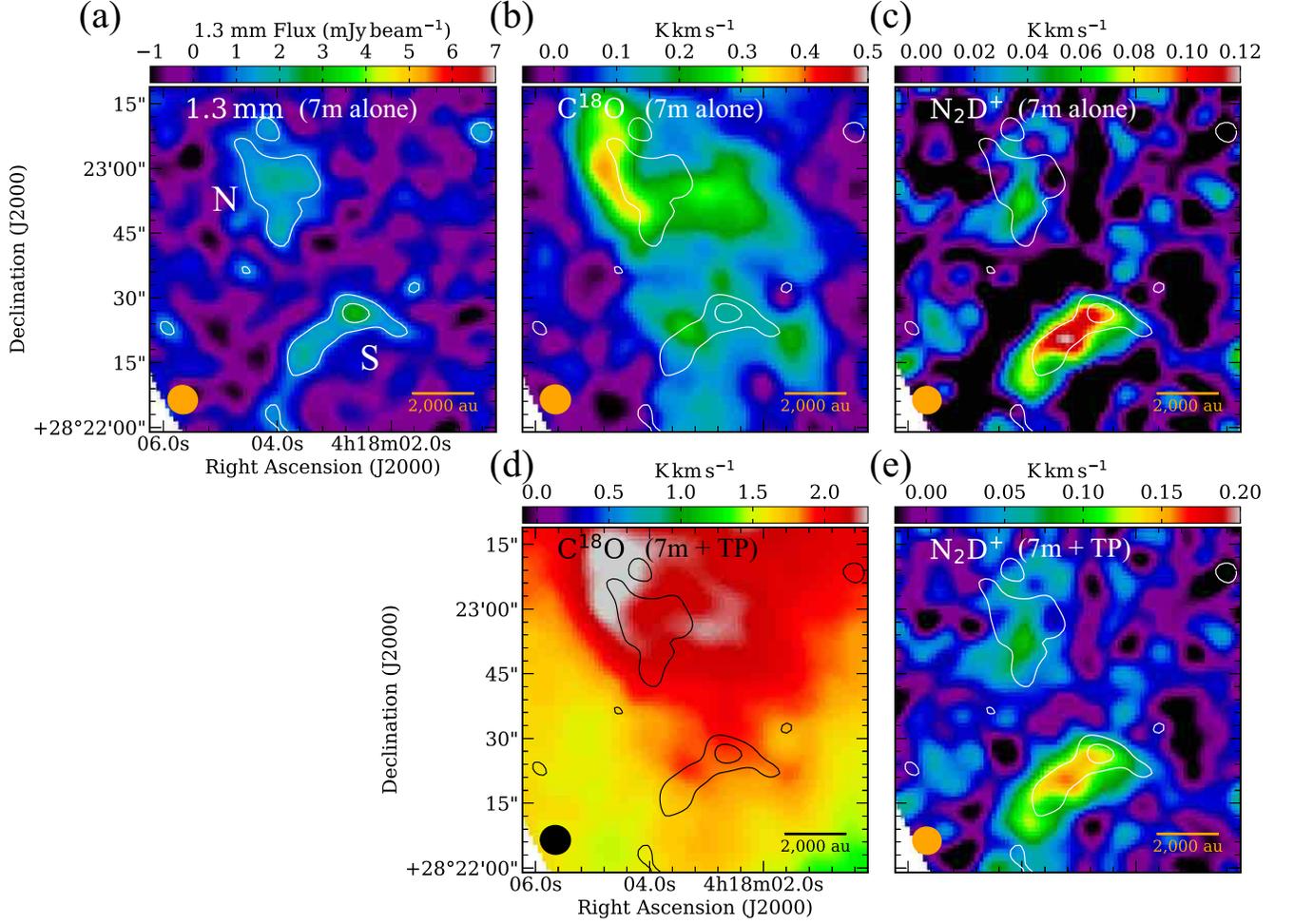}
\caption{Same as Figure\,\ref{fig:MC1_mol}, but for MC7N/S.  
\label{fig:MC7_mol}}
\end{figure}

Molecular line observations help us to evaluate the density distribution of dense cores. MC1 and MC7S/N have at least two significant peaks with an intensity of $\sim$3\,mJy\,beam$^{-1}$ in the 1.3\,mm continuum image (Figures\,\ref{fig:MC1_mol} (a) and \ref{fig:MC7_mol} (a)), although such a low-intensity contrast distribution can be reproduced by the interferometric artifact as discussed in the previous subsection. The molecular gas tracers (C$^{18}$O, and N$_2$D$^+$) tell us of further fruitful structures (Figures\,\ref{fig:MC1_mol} and \ref{fig:MC7_mol}). 

For MC1, the N$_2$D$^+$ distribution itself indicates that there is a density/size difference between the two continuum sources (Figure\,\ref{fig:MC1_mol} (c) and (e)). The southern source has an extended structure, while the size of the northern 1.3\,mm peak is close to the beam with a marginal detection in N$_2$D$^+$, indicating that the northern one is much smaller and less dense than the southern one. Another crucial evidence for the presence of density contrast is that the two continuum peaks sandwich the C$^{18}$O peak. Although this feature is more apparent in the 7\,m array image alone, which filters out the large-scale emission, the combined 7\,m + TP array image also reproduces the same trend (Figures\,\ref{fig:MC1_mol} (b) and (d)). If the innermost part of the dense core has a uniform density, but the density is high enough to detect N$_2$D$^+$, CO molecules are considered to be depleted on to dust grains and possibly show a ring-like structure surrounding the dusty central part \citep[e.g.,][]{Caselli02b,Crapsi05}. The present C$^{18}$O distribution in MC1 indicates that there is a relatively low-density ($\sim$10$^4$\,cm$^{-3}$) gas in between the overdense continuum peaks. 

Another promising candidate having significant substructures is MC7N/S. The molecular line distribution of C$^{18}$O and N$_2$D$^{+}$ follows a similar manner to that in MC1, as described above (see also Figure\,\ref{fig:MC7_mol}). The projected separation between 1.3\,mm peaks in each source is approximately 5000\,au, which is significantly larger than the flattening radius of the synthetic observation, producing fake substructures in the smoothed core (Figure\,\ref{fig:sim_obs}). In addition to this, the positions of each continuum source correspond to local peaks obtained by the early single-dish 0.87\,mm continuum observation (\citealt{Buckle15}, see also Figure\,\ref{fig:L1495} in Appendix\,\ref{dis3}). We thus exclude the interferometric artifact that produces the substructures over a few $\times$ 1000\,au in this case.
In summary, by considering the continuum and molecular line distributions, we confirmed that at least two prestellar sources harbor internal substructures, possibly produced by a fragmentation/coalescence process (see also \ref{dis:summary}). 

\subsection{Lifetime of 1000 au Scale Compact Structure within the Prestellar Cores in Taurus} \label{dis2}
We observed the center positions of $\sim$30 prestellar cores in Taurus. The observed targets are carefully selected based on the previous single-dish surveys in the dust continuum and molecular lines. The almost-complete sample allows us to statistically estimate the lifetime of high-density ($\sim$10$^{6}$\,cm$^{-3}$) peaks at the center of the core. \cite{Onishi02} discovered 44 prestellar cores with an average density of $\gtrsim$10$^{5}$\,cm$^{-3}$ and derived the timescale of protostar formation inside them, as $\sim$4\,$\times$\,10$^5$\,yr, assuming that the timescale of one of $\sim$100 pre–main-sequence stars in Taurus with ages of $\lesssim$10$^6$\,yr \citep{Kenyon90} is $\sim$10$^4$\,yr and their constant evolution speed.
They conclude that the lifetime of prestellar cores is several times longer than the freefall timescale of gas with 10$^{5}$\,cm$^{-3}$. This result is roughly consistent both with both mildly subcritical magnetized cores and with models invoking low levels of turbulent support \citep{Ward-Thompson07}. After their study, several observations additionally found similar density cores in molecular line observations \citep[e.g.,][]{Hirota02,Hacar13,Tafalla15,Arzoumanian18} 
in Taurus, and our survey includes two of them  (B10, L1521E). Our recent independent measurements in H$^{13}$CO$^+$, and N$_2$H$^+$ using the Nobeyama 45\,m telescope (K. Tokuda et al., in preparation) tell us that, in total, there are at most 10 prestellar cores, which are not presented in the \cite{Onishi02} catalog. Although we can possibly revise the total number of prestellar cores in Taurus and their lifetime to be 54, and $\sim$5\,$\times$\,10$^{5}$\,yr, respectively, the previous conclusion does not change that much.  

The present 7\,m array observations confirmed 1.3\,mm dust continuum emission toward 10 sources (see Sect.\,\ref{results:prestellar}). Note that if there are multiple local peaks from single observations (e.g., MC5, MC7, MC33), with at least one of them detected by the 7\,m array, we count it as one source. The total number of continuum-detected objects is significantly higher than that in the previous surveys, with zero or one successfully detected object by \cite{Dunham16} and \cite{Kirk17}. This discrepancy is a quite reasonable result because they targeted much higher density ($\gtrsim$10$^{7}$\,cm$^{-3}$) objects. They evaluated the detectability in continuum emission toward overdense regions within starless cores using the following equation,
\begin{equation}\label{eq:detect}
 {\rm Detection} > \frac{3}{2}\,\times\,N_{\rm total}\,\times\,\Bigl(\frac{n_{\rm Detectable}}{n_{\rm Limit}}\Bigr)^{-1}
\end{equation}
where $N_{\rm total}$ is the number of observed cores, $n_{\rm Detectable}$ is the central density threshold for detection, and $n_{\rm Limit}$ is the observed lower limit for the central number densities of the cores. With  $N_{\rm total}$ = 30, $n_{\rm Detectable}$ = $\sim$3\,$\times$\,10$^{5}$\,cm$^{-3}$ (Sect.\,\ref{dis1}), and $n_{\rm Limit}$ = $\sim$1\,$\times$\,10$^{5}$\,cm$^{-3}$, Equation\,\ref{eq:detect} tells us that the expected total number of detections is $\sim$11, which is consistent with our present observations.

\begin{figure}[htbp]
\begin{center}
\includegraphics[width=120mm]{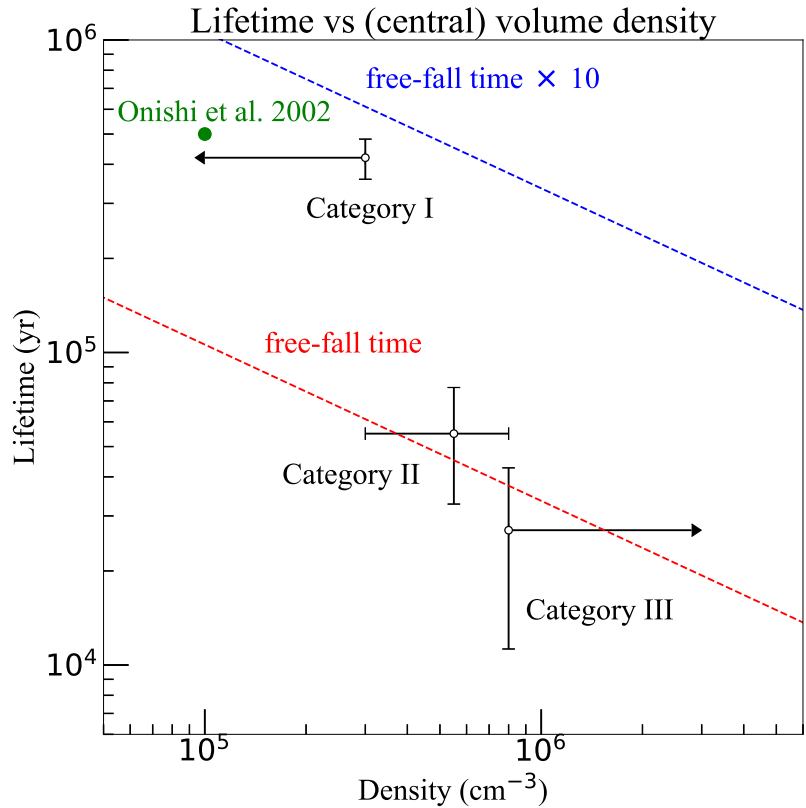}
\caption{Lifetime vs. (central) volume density. The X-axis error bars show the density range determined by comparisons between the real observations and synthetic ones (see the text in Sect.\,\ref{dis2}). The Y-axis error bar indicates $\sqrt{N}$ counting uncertainties. Note that the \cite{Onishi02} data point is the total lifetime that we assume to derive the lifetimes of each category.
\label{fig:densplot}}
\end{center}
\end{figure}

Our study gives us an observational constraint for the lifetime of the innermost parts of prestellar cores on the verge of star formation. Based on our 7\,m array 1.2/1.3\,mm continuum measurements, we divided the 54 prestellar sources, whose lifetime is $\sim$5\,$\times$\,10$^{5}$\,yr, in Taurus (see the first paragraph in this subsection) into three categories: (I) cores without continuum, (II) cores with weak continuum, and (III) cores with strong continuum (see Table\,\ref{table:1mm_pre}). We tentatively set a flux criterion of 21\,mJy, whose density is $\gtrsim$8\,$\times$10$^{5}$\,cm$^{-3}$, to separate II and III. The total numbers in categories I, II, and III are 45, 6, and 3, respectively. We assumed that the unobserved cores in this study are categorized under I, because even the single-dish observations indicate a less evolved feature than our selected target (see the Introduction). If a single core has two peaks with different categories (e.g., MC7N/S and MC33bS/N), we counted the core as the latter stage.
The central density ranges of each category are roughly $\lesssim$3\,$\times$\,10$^5$\,cm$^{-3}$ (I), (3--8)\,$\times$\,10$^5$\,cm$^{-3}$ (II), and $\gtrsim$8\,$\times$\,10$^5$\,cm$^{-3}$ (III). If we adopt the lifetime of prestellar cores, $\sim$5\,$\times$\,10$^{5}$ yr, the timescale of each stage can be divided into $\sim$3\,$\times$\,10$^{5}$\,yr (I), $\sim$6\,$\times$\,10$^{4}$\,yr (II), and $\sim$3\,$\times$\,10$^{4}$\,yr (III). We exhibit the estimated lifetimes in Figure\,\ref{fig:densplot}, which is similar to the $``$JWT plot$"$ (after \citealt{Jessop00}, see also Figure\,2 in \citealt{Ward-Thompson07}) with a density range of 10$^3$--10$^{5}$\,cm$^{-3}$. For the less dense cores (category I), the lifetime is several or a few times longer than the freefall time of the gas density. This means that supporting forces, such as the magnetic field and turbulence, still play a role in preventing the freefall collapse. On the other hand, in the further high-density regime, the timescale approaches the freefall time (see also \citealt{Konyves15}). We thus propose that the density threshold is $\sim$10$^6$\,cm$^{-3}$ when self-gravity becomes the dominant force regulating the core dynamics. Infalling motions may be detectable in the latest stage. However, if the collapsing radius is smaller than the beam sizes of single-dish telescopes, detecting blue asymmetric profile emission is difficult (see the discussion for MC5N, one of the category III objects, by \citealt{Tokuda19}). Interferometric observations with high-$J$ transitions of optically thick tracers are possible methods to further confirm the internal core dynamics (see also Sect.\,\ref{sec:future}).

We briefly mention the lifetime of the possible candidate for the FHSC in MC35. From the dust peak, we found a bipolar outflow, whose dynamical age is less than 10$^{4}$\,yr. Paper II described that the observed nature of MC35 to be consistent with the theoretical properties of the FHSC \citep[e.g.,][]{Machida08}.
Finding only one candidate out of the dozen starless targets indicates that the lifetime is less than 10$^4$\,yr (= 5\,$\times$\,10$^5$\,yr * 1/54). This is marginally consistent with that from theoretical predictions, $\sim$10$^{3}$\,yr \citep[e.g.,][]{Larson69}. We further discussed the lifetime of the FHSC candidate and the comparison between our observations and some numerical simulations in Paper II.

\subsection{Dense Core Evolution and Its Substructure Formation in the Taurus Molecular Cloud}\label{dis:summary}
Most of the dense cores, including our present targets in Taurus, line in filamentary clouds, suggesting that dense core formation is deeply related to the kinematics of filamentary gas, such as fragmentation and collapse as suggested by the previous studies (\citealt{Onishi98,Onishi02}, see the general review by \citealt{Andre14}).  
Our present study features the subsequent fate after the dense core formation with a $\sim$1000\,au spatial resolution.

Early single-dish observations revealed the diversity of shapes of prestellar sources: for example, nearly spherical, prolate/oblate structure, and highly irregular morphology based on the dust continuum and molecular line observations \citep[e.g.,][]{Onishi02,Caselli02b,Kauffmann08}. The large-scale structures of category III objects have relatively axisymmetric shapes in their strong intensity levels (e.g., more than the FWHM contour; see \citealt{Tokuda19} for MC5N/S, and \citealt{Kauffmann08} for MC31). This fact is consistent with our suggestion that self-gravity is eventually responsible for the dynamics of prestellar cores whose central density exceeds more than 10$^6$\,cm$^{-3}$. 

For category II, some of the parental cores show a highly irregular shape \citep[e.g., MC33, see][]{Onishi02,Caselli02a,Kauffmann08}. In this stage, in addition than self-gravity, magnetic field and turbulence may also play an essential role in forming high-density maxima, which are detectable with interferometers. 
In addition to this, our measurements discovered at least two promising candidates whose internal substructures have a size scale of $\sim$1000\,au. Although their actual origin and whether or not the substructures may harbor or go on to form protostars individually remain to be studied, the new findings suggest that the molecular cloud cores are not necessarily isolated objects, and their mass has a time dependence during the fragmentation/coalescence processes.

Intriguingly, the cores with the possible substructures belong to category II. We suggest that substructure formation may occur before the self-gravity controls their fate. Theoretical studies do not prohibit coalescence among multiple cores after the fragmentation of their parental filamentary cloud, depending on the actual perturbation in the filament \citep[c.f.,][]{Inutsuka97,Masunaga99,Inutsuka01}. If we assume an initial core separation of $\sim$0.1\,pc and the velocity of the core motion is similar to the isothermal sound speed, $\sim$0.2\,km\,s$^{-1}$, for a gas temperature of 10\,K, the expected time scale of the coalescence is $\sim$5\,$\times$\,10$^5$\,yr (=0.1\,pc/0.2\,km\,s$^{-1}$). This time scale is close to the statistical lifetime of the Taurus dense core with a density of $\sim$10$^5$\,cm$^{-3}$ (see Sect.\,\ref{dis2}), indicating that coalescence of dense cores can occur within their lifetime.  

Regarding fragmentation processes, \cite{Nakamura97} demonstrated that the core-forming clouds become unstable to a bar mode after the magnetically supercritical core formation \citep[see also][]{Machida05,Nakamura02}. The two overdense regions in MC7 are distributed along the major axis of the parental core (see Figure\,\ref{fig:L1495} in Appendix\,\ref{dis3}). Although bar-mode fragmentation is plausible in this case, MC1 shows the opposite trend (see the single-dish continuum observation by \citealt{Motte01}). These diversities imply that there are several mechanisms that promote fragmentation/coalescence processes.

\section{Summary} \label{sec:sum}

We have carried out a survey-type project toward 32 prestellar and 7 protostellar cores in the Taurus main filamentary complex using the ALMA-ACA (Atacama Compact Array, the 7\,m + TP array) stand-alone mode with an angular resolution of 6\farcs5 ($\sim$900\,au). Our main conclusions can be summarized as follows: 
\begin{enumerate}
\item A large fraction (30--90\%) of the continuum emission from the protostellar cores are contributed by protostellar disks, except in the very low-luminosity protostar case.
The continuum observations toward the prestellar sources have revealed the presence/absence of a compact inner structure at the center whose detection rate is approximately one-third. Thanks to the lower spatial frequency coverage with the 7\,m array, the success rate is significantly higher than the previous ALMA main array surveys. The continuum-detected prestellar cores have a central density, $n_{\rm c}$, of $\gtrsim$3\,$\times$\,10$^5$\,cm$^{-3}$ and are more evolved than the remaining sources without continuum detection (category I, $n_{\rm c}$ $\lesssim$3\,$\times$\,10$^{5}$\,cm$^{-3}$).

\item Statistical counting of the continuum-detected sources tell us the lifetime of such a high central density object. The subsample of weak continuum-detected sources (category II, $\sim$(3--8)\,$\times$\,10$^{5}$\,cm$^{-3}$) shows that its lifetime is slightly longer than the freefall time of the gas density, while the prestellar cores with strong continuum emission (category III, $\gtrsim$8\,$\times$\,10$^{5}$\,cm$^{-3}$) have a much shorter timescale, which is close to the freefall time. This result suggests that the threshold density to dominate the core dynamics by self-gravity during dense core evolution is $\sim$10$^6$\,cm$^{-3}$.

\item Some of the continuum-detected prestellar sources have complex substructures with the size scale of $\sim$1000\,au. Synthetic observations with the 7\,m array toward smoothed core models with a flattening radius of a few $\times$ 1000\,au can mimic such a structure due to the interferometric effect \citep[see also][]{Caselli19}.
However, molecular line observations in C$^{18}$O and N$_2$D$^+$ indicate that there is a real density contrast between each continuum peak in two category II objects, MC1 and MC7. 
The presence of substructures with a size scale of $\sim$1000\,au suggests that dense cores are not necessarily isolated objects, and small-scale fragmentation/coalescence processes within the $\sim$0.1\,pc core that affect the final mass likely happen before the dynamical collapse to form stars.
\end{enumerate}

\section{Future prospects} \label{sec:future}
This work was the first comprehensive dense core survey with the ACA stand-alone mode toward a low-mass star-forming molecular cloud complex and provided further motivations for branching out into several follow-up studies. (1) High-resolution observations using the ALMA main (12\,m) array toward some possible candidates just before/after star formation (e.g., MC5N, MC35) will elucidate the precise nature of fragmentation and collapse over a few hundred astronomical unit scale or less. (2) The present line setting alone cannot fully explore the gas properties, and thus an alternative frequency setup is also needed to understand their kinematics. For example, optically thick tracers with a high critical density, such as HCO$^+$ and CS, are useful to trace the collapsing motion of dense cores by looking at their blue asymmetric profile \citep[e.g.,][]{Lee04}. (3) Although we have to consider the distribution of the magnetic field to understand the stability of the dense cores, the current ACA capability does not allow us to perform polarization observations. Until we get the function, single-dish observations by, e.g., JCMT, IRAM, and SOFIA will play an important role in obtaining the polarized emission with a relatively low spatial frequency component of dense cores.

Extending this type of survey toward other molecular clouds is also crucial to obtain the general picture of star formation. High-density cores just before star formation are quite rare and difficult to find, as statistically demonstrated by the early studies as well as our current project. Nearby ($D\sim$150\,pc) low-mass star-forming regions, such as Lupus and Ophiuchus-North, are promising candidates to be observed with the ACA. R CrA, $\rho$ Ophiuchus (see also, \citealt{Kamazaki19}), and B59 in the Pipe Nebula are also vital targets to compare the properties of dense cores between the isolated low-mass star-forming regions and low-to-intermediate cluster-forming sites, and we will eventually study the environmental effects on the dense core evolution.

\acknowledgments
This paper makes use of the following ALMA data: ADS/ JAO.ALMA\#2015.1.00340.S, 
2016.1.00928.S, and 2018.1.00756.S. ALMA is a partnership of ESO (representing its member states), NSF (USA) and NINS (Japan), together with NRC (Canada), MOST and ASIAA (Taiwan), and KASI (Republic of Korea), in cooperation with the Republic of Chile. The Joint ALMA Observatory is operated by ESO, AUI/NRAO, and NAOJ. This work was supported by NAOJ ALMA Scientific Research grant Nos. 2016-03B and Grants-in-Aid for Scientific Research (KAKENHI) of Japan Society for the Promotion of Science (JSPS; grant Nos. 18K13582, and 18H05440). We thank Dr. Doris Arzoumanian and Dr. Pedro Palmeirim for discussions on the filamentary clouds and prestellar cores in Taurus. Dr. Paul F. Goldsmith kindly provided us with a $^{12}$CO\,($J$ = 1--0) data cube obtained by the FCRAO survey \citep{Goldsmith08,Narayanan08} to search for CO emission-free positions as reference (OFF) points for the TP array observations.
\software{CASA (v5.6.0; \citealt{McMullin07}), Astropy \citep{Astropy18}, APLpy (v1.1.1: \citealt{Robi12}}

\appendix

\section{The data reduction and image qualities} \label{AppA}

\subsection{1.3 mm Continuum Imaging with Two Different Spectral Windows}\label{App:Two_cont}
\begin{figure}[htbp]
\includegraphics[width=180mm]{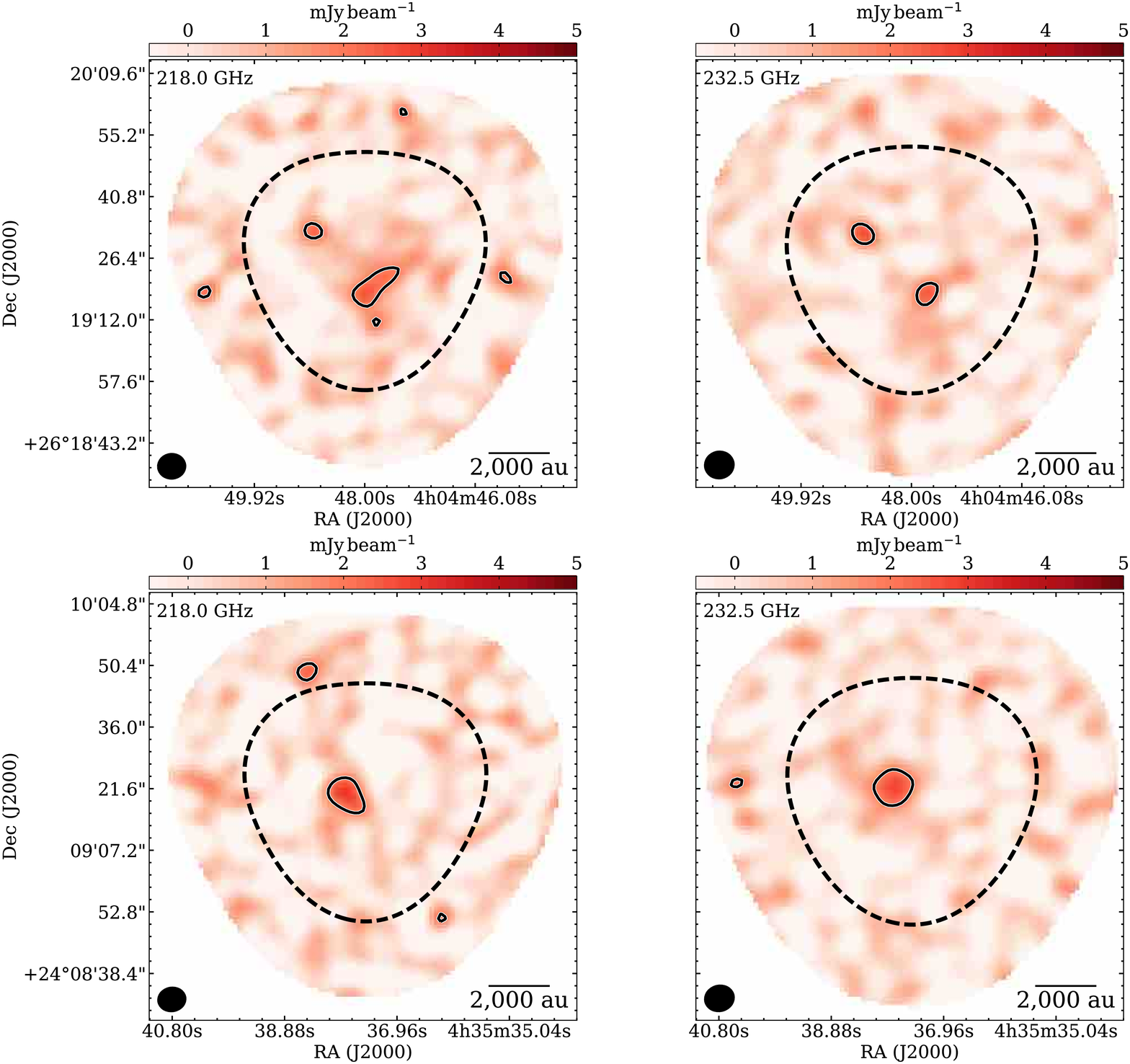}
\caption{218.0\,GHz and 232.5\,GHz continuum distributions of MC1 (upper panels) and MC35 (lower panels). Dashed black lines indicate where the mosaic sensitivity falls to 50\%. The contour level is 3$\sigma$ noise level ($\sim$0.6\,mJy\,beam$^{-1}$) of each measurements. The lower-left corners give the angular resolutions. Note that the primary beam attenuation is not corrected for the display purposes. 
\label{fig:cont1-2}}
\end{figure}

We made continuum images with two individual spectral windows whose central frequencies are 218.0 and 232.5\,GHz (see Sect.\,\ref{obs:freq}). As representative examples, we performed the imaging of MC1, which is one of the promising candidates with internal substructures (see Sect.\,\ref{dis1}), and MC35, the first core candidate (see Paper II). Figure\,\ref{fig:cont1-2} shows that the two frequency images reproduce almost the same result.  

\subsection{The Image Qualities}
Table\,\ref{table:beam_rms} gives the beam sizes and rms sensitivities of each continuum map. The average angular resolution is 6\farcs8\,$\times$\,6\farcs2, and the geometric mean is 6\farcs5. We quote this value as the representative angular resolution in the abstract and summary.

\begin{table}[htbp]
\caption{Beam Properties and Sensitivities in the Continuum Observations}
\label{table:beam_rms}
\centering
\begin{tabular}{lcccc} \hline
Name & $B_{\rm maj}$\,(arcsec.) & $B_{\rm min}$\,(arcsec.) & $B_{\rm P.A.}$ (deg.) & rms\,(mJy\,beam$^{-1}$) \\ \hline
MC1  & 6.8  & 6.3  & -87.9   & 0.37  \\
MC2  & 6.8  & 6.1  & 84.7  & 0.40  \\
MC4  & 6.9  & 6.5  & 78.2  & 0.39  \\
MC5N  & 6.9  & 6.5  & 67.4  & 0.38  \\
MC5S & 6.9  & 6.5  & 69.3  & 0.38  \\
MC6  & 7.5  & 5.2  & -54.9  & 0.55 \\
MC7N & 6.9  & 6.5  & 71.7  & 0.37  \\
MC7S & 6.8  & 6.5  & 71.2  & 0.38  \\
MC8  & 7.5 & 5.2 & -54.9 & 0.65 \\
MC11 & 6.8  & 6.5  & 75.8  & 0.40  \\
B10  & 6.8  & 6.5  & 83.6  & 0.38  \\
MC13W  & 6.8  & 6.4  & 83.8  & 0.36  \\
MC13a  & 6.8  & 6.4  & 85.7  & 0.38  \\
MC13b  & 6.9  & 6.4  & 87.7  & $\cdots^{\rm a}$ \\
MC14N  & 6.8  & 6.4  & -90.0   & 0.43  \\
MC14S  & 6.9  & 6.4  & -88.6   & $\cdots^{\rm a}$ \\
MC16E  & 6.8  & 6.4  & -81.1   & 0.33  \\
MC16W  & 6.8  & 6.4  & -78.1   & 0.41  \\
MC19 & 6.8  & 6.1  & -86.6   & 0.38  \\
MC22 & 6.8  & 6.3  & -82.4   & 0.36  \\
MC23 & 6.8  & 6.1  & -86.7   & 0.35  \\
MC24 & 6.8  & 6.1  & -86.3   & 0.42  \\
MC25E  & 6.8  & 6.3  & -84.1   & 0.46  \\
MC25W  & 6.8  & 6.3  & -77.1   & 0.41  \\
MC26a  & 6.8  & 6.3  & -77.1   & 0.37 \\
MC27 & 7.2  & 5.8  & -59.6  & 0.30 \\
MC28 & 6.9  & 6.1  & -83.8   & 0.44 \\
MC29 & 6.8  & 6.0  & -80.7   & 0.45  \\
MC31 & 6.8  & 6.0  & -81.2   & 0.40  \\
MC33bS  & 6.8  & 6.0  & -81.2   & 0.33  \\
MC33bN  & 6.8  & 6.0  & -80.0   & 0.45  \\
MC34 & 6.9  & 5.9  & -82.1   & 0.42  \\
MC35 & 6.8  & 6.0  & -80.8   & 0.39  \\
MC37 & 6.8  & 6.2  & -77.4   & 0.35  \\
MC38 & 6.8  & 6.2  & -76.6   & 0.31  \\
MC39 & 6.9  & 6.2  & -77.6   & $\cdots^{\rm a}$ \\
MC41 & 6.8  & 6.2  & -74.9   & $\cdots^{\rm a}$ \\
MC44 & 7.3  & 5.2  & -50.1   & 0.61 \\
L1521E & 7.0  & 6.1  & -63.3   & 0.39 \\ \hline
\end{tabular}
\tablenotetext{\rm a}{The strong peak intensities make it difficult to accurately measure the sensitivities due to the sidelobe effect. For these sources, we apply the typical sensitivity, 0.4\,mJy\,beam$^{-1}$, in PROJ6 to draw the contours in Figures\,\ref{fig:all_core} and \ref{fig:B213}.}
\end{table}

\subsection{Comparison between the Real and Synthetic Observations} \label{Ap:Obs_model}
\begin{figure}[htbp]
\includegraphics[width=180mm]{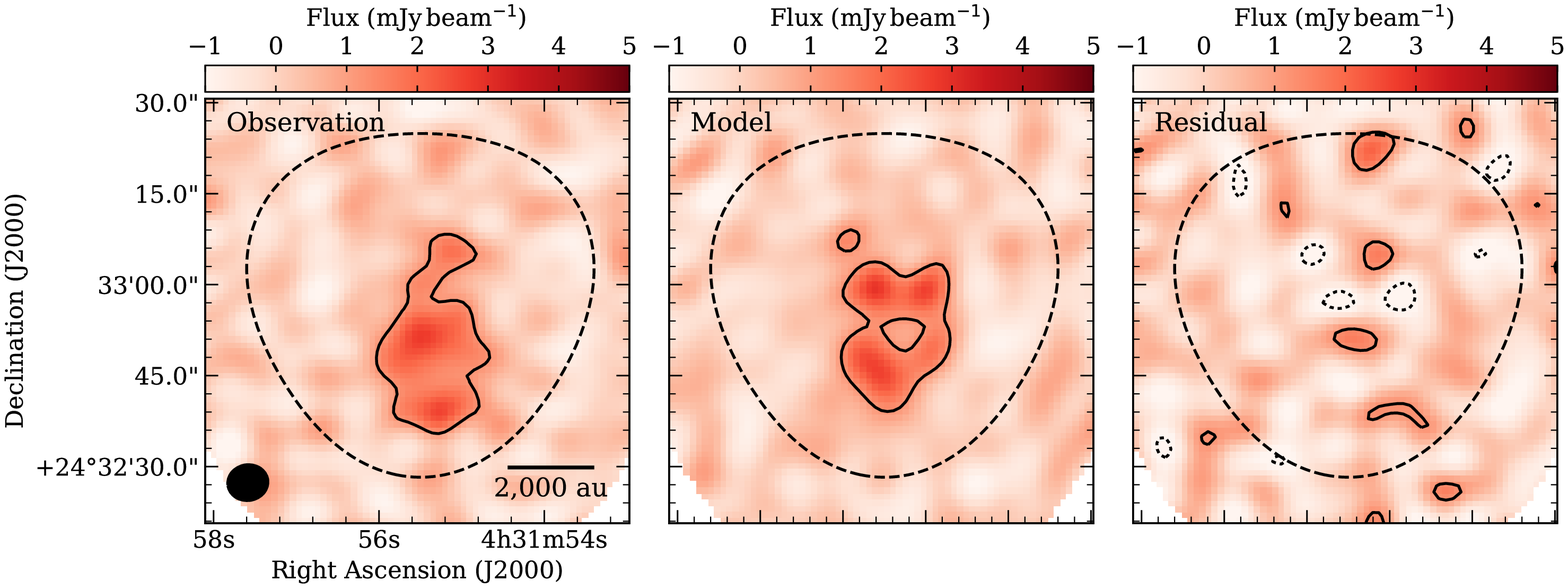}
\caption{Comparison between the real and synthetic observations in 1.3\,mm continuum with the 7\,m array. The black ellipse in the lower-left corner gives the observed angular resolution. The left panel shows the real observations of MC31, which is the same as that in Figure\,\ref{fig:all_core}. The middle panel is the synthetic observation of the smoothed core model (see the text). The right panel represents the residual image obtained by subtracting the model image from the observed one. The solid and dotted contours in each panel are the positive and negative 3$\sigma$ levels, respectively. The dashed lines indicate where the mosaic sensitivity falls to 50\%.
\label{fig:residual}}
\end{figure}

Figure\,\ref{fig:residual} illustrates the comparison between the real and synthetic observations of MC31 as a representative example. We adopted an aspect ratio of 1.8 and a position angle of 4\arcdeg\ estimated from the real observation (Table\,\ref{table:1mm_pre}) as input parameters of the parametric model. The synthetic core with an $R_{\rm flat}$ of 2500\,au and $N_{\rm H_2}$ ($r$ = 6000\,au) = 1\,$\times$10$^{22}$\,cm$^{-2}$ reproduces total and peak fluxes similar to the real observation. The residual image does not show large differences between the real and synthetic observations. This result demonstrates that the synthetic model is reasonable to fit the observed image and that interferometric observations can artificially produce fake substructures even if the core has a smooth distribution in nature.

\section{Clustered Cores at Individual Subregions in Taurus} \label{dis3}
In the Taurus main filamentary complex (Figure 1), there are several regions with clustering cores. We discuss their properties and evolutionary stages by combining early studies and our new measurements.

\subsection{The B213 Region} \label{dis:B213}
\begin{figure}[htbp]
\includegraphics[width=180mm]{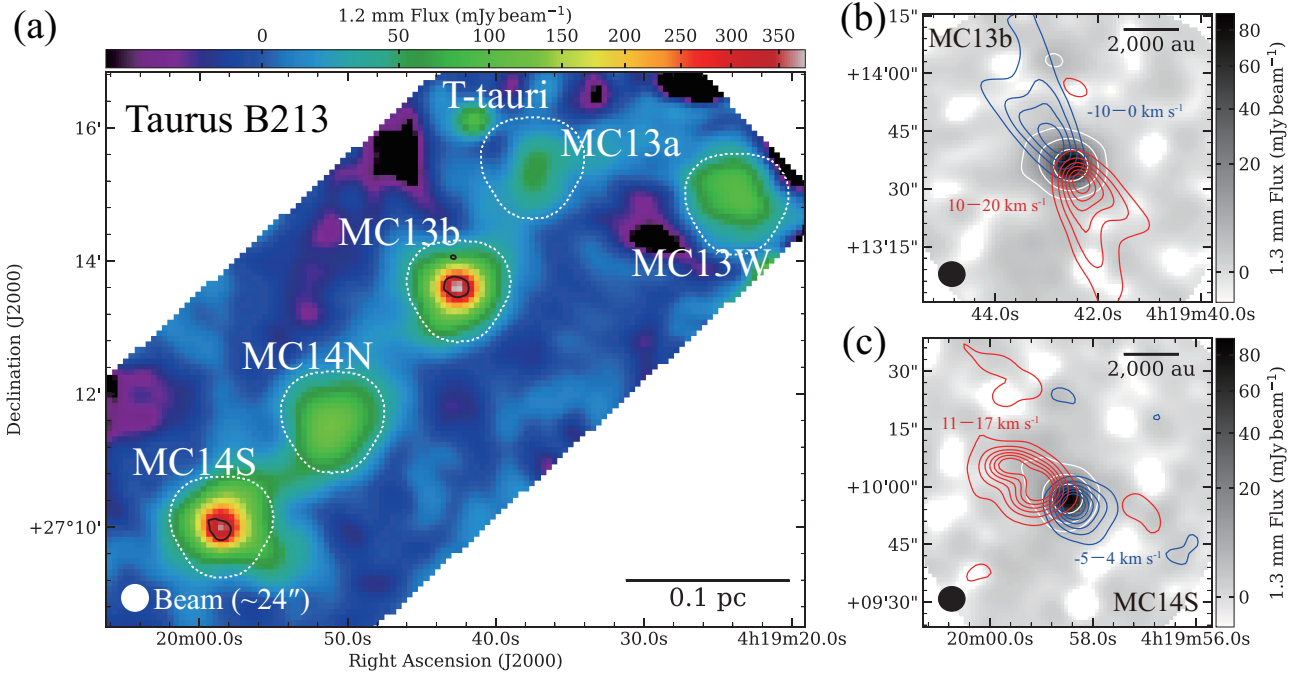}
\caption{Dust continuum and $^{12}$CO outflow distributions toward the Taurus B213 region. (a) The 1.2\,mm continuum obtained by the IRAM\,30\,m/NIKA2 \citep{Bracco17}. The beam size, $\sim$24\arcsec, is shown by the white circle in the lower-left corner. White dotted lines show the field coverage of the 7\,m array observations. The black contours are the 7\,m array continuum in 1.3\,mm with a contour level of 10$\sigma$. (b) and (c) Gray-scale maps show the 7\,m array observations in 1.2\,mm continuum toward MC14S and MC13b. Red and blue contours show redshifted/blueshifted $^{12}$CO (2--1) emission obtained with the 7\,m array. The integrated velocity ranges are given at the vicinities of each minimum contour in the figures. The lowest and subsequent contour steps are 0.4 and 1.2\,K\,km\,s$^{-1}$, respectively. The angular resolution, 6\farcs8\,$\times$\,6\farcs5, is given by black ellipses in each panel. The white contour shows the 1.2\,mm continuum with the contour levels of 10$\sigma$, 30$\sigma$, and 100$\sigma$.
\label{fig:B213}}
\end{figure}

Figures\,\ref{fig:B213} shows 1.2/1.3\,mm continuum distributions toward the B213 filamentary cloud. We observed three prestellar and two protostellar cores in this region. The line mass of the filament is larger than the critical line mass, $M_{\rm line,crit}$ = 2$c_{\rm s}^2$/G \citep[e.g.,][]{Stodolkiewicz63,Ostriker64,Inutsuka92}, where $c_{\rm s}$ $\sim$0.2\,km\,s$^{-1}$ is the isothermal sound speed for a gas temperature of $\sim$10\,K. It means that the filament is unstable against the fragmentation and collapse. In this region, several prestellar and prestellar cores are alternate with a separation of $\sim$0.1\,pc (e.g., \citealt{Onishi02,Hacar13} see also Figure\,\ref{fig:B213}). There is no other star-forming filament showing such a regular distribution in Taurus. The present 7\,m array observations could not find the 1.3\,mm continuum emission toward all of the prestellar sources in the B213 filament (Figure\,\ref{fig:B213} (a)), suggesting that they are not sufficiently centrally concentrated. The onset of dynamical collapse is one option to get that configuration, i.e. detection of continuum emission with the 7\,m array.

Previous studies suggested that a large-scale colliding accretion flow created the B213 filament \citep{Palmeirim13,Shimajiri19}. If the filament initially has a uniform density with a 0.1\,pc width, which is a quasi-universal value in nearby molecular clouds \citep{Arzoumanian11,Arzoumanian19}, we cannot explain such an evolutionary difference unless there was a density fluctuation in this system at the formation phase. In MC14S and MC13b, the fluctuation of the filament formed overdense regions, and then it might have collapsed into the protostar faster than the other cores. In this case, the alternating distribution of pre-/protostellar cores itself is a coincidence rather than having some inherent physical meanings.

The present frequency setting allows us to investigate the outflow distribution in $^{12}$CO. Panels (b) and (c) in Figure\,\ref{fig:B213} show the directions of the $^{12}$CO outflows, which are consistent with the early interferometric measurements (\citealt{Takakuwa18}, for MC14; \citealt{Tafalla16}, for MC13b). 
The outflow axis seems to be perpendicular to the parental filament elongation, but not in the case of MC14S. This feature means that the large-scale kinematics not only determines the rotation axis of the protostellar disk that originated from the filament fragmentation, but also local phenomena within the dense core. \cite{Takakuwa18} found a counterrotation between the disk and protostellar envelope in MC14S. They interpreted that the magnetic field may affect the formation of such a complex system.

\subsection{The L1495 Region} \label{dis:L1495}
\begin{figure}[htbp]
\begin{center}
\includegraphics[width=160mm]{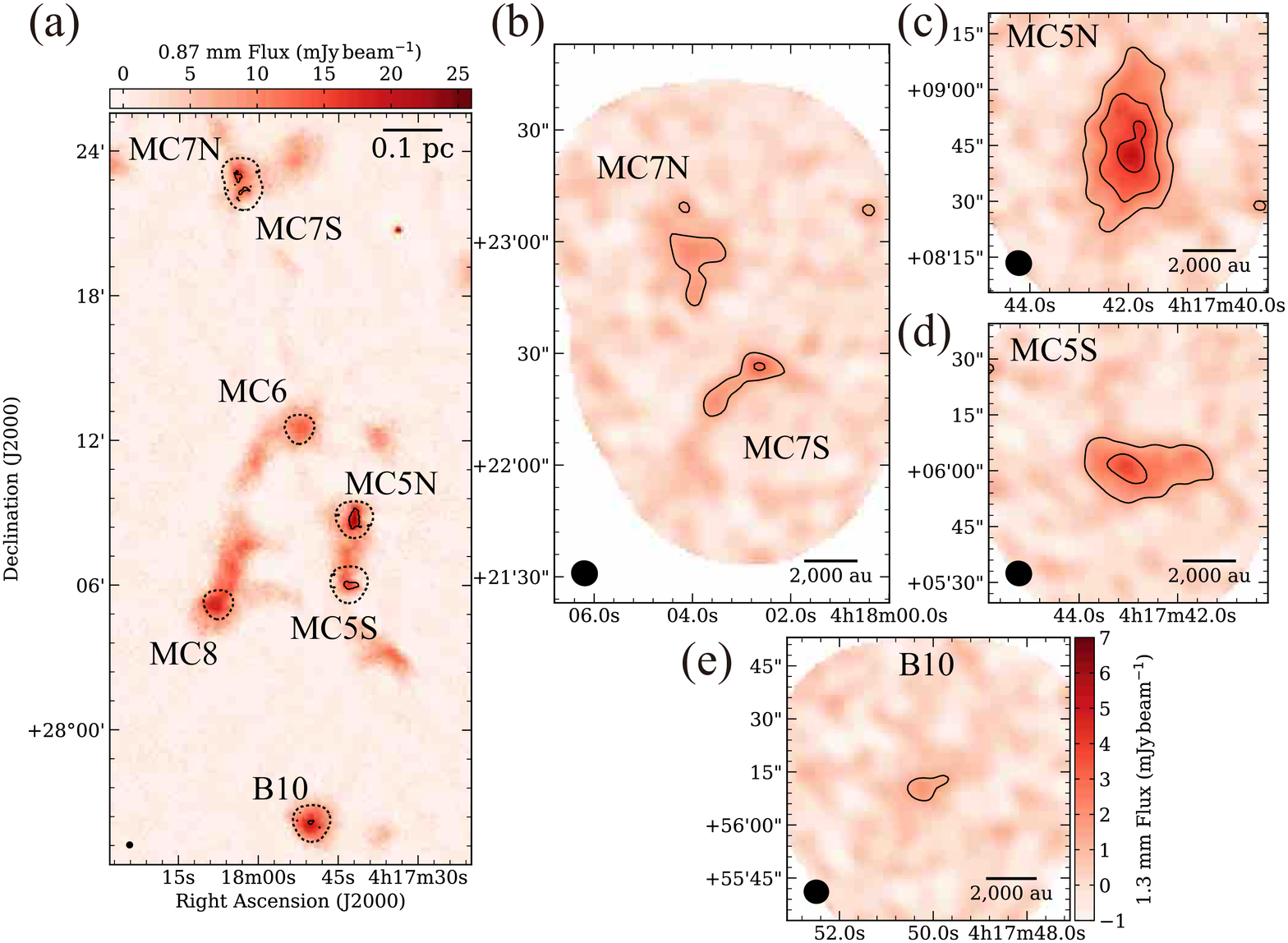}
\caption{0.87/1.3\,mm continuum images toward the L1495 region. (a) Color-scale image shows the 0.87\,mm continuum map obtained by JCMT/SCUBA-2 \citep{Buckle15}. White dotted lines show the field coverages of our 7\,m array observations. Black contours show the 7\,m array continuum images at the $\sim$3$\sigma$ level. (b--e) Color-scale images and contours show 1.3\,mm continuum emission observed by the 7\,m array. Ellipses in each lower-left corner show the beam sizes.  
\label{fig:L1495}}
\end{center}
\end{figure}

A previous single-dish study found a remarkable filamentary complex in the L1495 region. We observed seven prestellar cores in Figure \ref{fig:L1495} (a). In this region, more than 10 dense cores are clustering without any indication of star formation. 
As shown by early molecular line observations in $^{13}$CO and C$^{18}$O \citep{Mizuno95,Hacar13}, several filamentary structures are entangled with each other toward this region, indicating that the surrounding filamentary gas accreted onto the primary dense filaments and then they fragmented into several cores \citep{Tafalla15}. Our 7\,m array continuum observations found internal structures toward the prestellar cores except for MC6 and MC8, suggesting that these cores may be highly evolved and collapse into protostars soon.

\cite{Tokuda19} reported that MC5N has a high-density ($\sim$10$^6$\,cm$^{-3}$) peak at the center of the core, and it is a promising candidate of a brown dwarf prestellar core based on its small parental core mass, $\sim$0.2--0.4\,$M_{\odot}$. We predicted that another subcore, MC5S, also has a similar density enhancement, if they were formed by a common mechanism, which is the fragmentation of a filamentary cloud after the radial collapse \citep{Inutsuka92,Inutsuka97}. The detection of 1.3\,mm continuum and N$_2$D$^+$ in MC5S indicate that there is a high-density peak, and it is consistent with our prediction. Although the early single-dish study already confirmed the two local peaks of the core \citep{Buckle15,Ward-Thompson16}, we also found 1.3 mm continuum detection with the 7\,m array at the positions of MC7N and MC7S. A crucial difference between MC5 and MC7 is the length of each separation within their system; the projected distance of two subcores in MC5 is as long as $\sim$0.1 pc, while that in MC7 is much shorter, $\sim$0.02\,pc. 

\subsection{The B18 region} \label{dis:B18}
The B18 cloud is at the southern part of the Taurus (Figure \ref{fig:Taurusmap}), and the star formation is more active there than at the western side, the L1531 region \citep[e.g.,][]{Mizuno95}. Although the overall distribution roughly shows a filamentary structure, the individual dense cores have highly complex morphologies. The separation between the cores is sparse compared to that in the other subregion, e.g., B213/L1495. The recent survey using the Green Bank Telescope \citep{Friesen17} detected the ammonia emission toward the bright regions in the Herschel dust continuum observations (Figure\,\ref{fig:Taurusmap}). There are only three prestellar sources without the 1.3\,mm continuum emission from the 7\,m observation. The detection rate is similar to that in L1495. 

In this region, there are some intriguing sources in terms of the early phases of star formation. As mentioned in Sect.\,\ref{results:prestellar}, the MC35 is a possible candidate for the FHSC, because the source has no bright infrared sources and the CO observations found a possible compact bipolar outflow. In MC24, we also detect a redshifted high-velocity wing. Because the high-velocity component is not connected to the dust continuum peak, we cannot exclude the possibility that the high-velocity component is part of the large-scale outflow from a nearby protostellar source, IRAS\,04239+2436 \citep{Narayanan12}. We thus suppose that the probability of a protostellar object being contained in MC24 seems to be lower than that of MC35. 

We observed two positions in MC33; one has a dust continuum peak (MC33bS) in the single-dish observation \citep{Kauffmann08}, and the other shows the N$_2$H$^+$, N$_2$D$^{+}$, and NH$_3$ peak \citep{Caselli02a,Crapsi05}. If we accept that the N-bearing species are tracers of chemically evolved regions, the current observations suggest that MC33bS is a physically evolved peak with a high column density, while MC33bN is a more chemically evolved part. Our 7\,m array observations also detected 1.3\,mm continuum emission in MC33bS, not MC33bN. This result is consistent with the previous studies. MC33 is a vital target for examining the inhomogeneity between a mass distribution and its chemical composition.

MC28 is an interesting target harboring both the protostellar source IRAS 04263+2426 and two starless peaks. As mentioned in Sect.\,\ref{results:protostellar}, they share almost the same systemic velocity. This fact indicates that the fragmentation process of the parental core or the protostellar feedback produced the dense starless blobs. This system can be an excellent candidate to study the formation of a wide binary/multiple. 

\subsection{The HCL2 region} \label{dis:HCL2}
The HCL2 (Heiles cloud 2) region \citep{Heiles68} has ring-like or filamentary structures observed by dense gas tracers and dust continuum emission (e.g., \citealt{Onishi96,Feh16}, see also Figure\,\ref{fig:Taurusmap}). \cite{Feh16} estimated the gas temperature from the NH$_3$ observations. The Mt. Fuji telescope found a C$\;${\sc i} peak, whose position is away from the dense material, in the eastern part of the region \citep{Maezawa99}. This result indicates that the region is a chemically young region where C$\;${\sc i} has not yet been fully converted into CO and the evolutionary sequence propagates from the eastern to western side. The absence of the 1.3\,mm continuum emission in MC44 in our measurement means that the core is in a relatively young phase before the dynamical collapse and follows the global evolutionary sequence of this region. On the western side, there are several famous star-forming cores, such as L1527 and MC39(=TMC-1A). As shown in Table\,\ref{table:1mm_proto}, the $F_{\rm disk}$/$F_{\nu}$ in MC39 is higher than that in other cores harboring class 0/I protostars (MC13b, MC14S). This result indicates that the extended gas in MC39 has been accreted onto the protostar and protostellar disk, which is consistent with the prediction that the protostar is in a late class I phase \citep{Aso15}.

MC37 is a prestellar core with weak continuum emission in our survey, although the complex distribution itself is possibly arises from the interferometric artifact (see Sect.\,\ref{dis2}).

\bibliography{sample63}{}
\bibliographystyle{aasjournal}



\end{document}